\documentclass{aa}  
\usepackage{graphicx}
\usepackage{txfonts}
\usepackage{amsmath}
\usepackage{comment}
\usepackage{enumitem}
\usepackage{xcolor}
\usepackage{placeins}
\usepackage{multirow}
\usepackage{caption}

\begin{document}

   \title{Tracing the Milky Way warp and spiral arms\\ with classical Cepheids}

   %\subtitle{}

   \author{B. Lemasle
          \inst{1}
          \and
          H. N. Lala
          \inst{1}
          \and
          V. Kovtyukh
          \inst{2}
          \and
          M. Hanke
          \inst{1}
          \and
          Z. Prudil
          \inst{1}
          \and
          G. Bono
          \inst{3,4}
          \and
          V. F. Braga
          \inst{4,5}
          \and
          R. da Silva
          \inst{4,5}
          \and
          M. Fabrizio
          \inst{4,5}
          \and
          G. Fiorentino
          \inst{4}
          \and
          P. Fran\c cois
          \inst{6,7}
          \and
          E. K. Grebel
          \inst{1}
          \and
          A. Kniazev
          \inst{8,9,10}
          }

   \institute{Astronomisches Rechen-Institut, Zentrum f\"ur Astronomie der Universit\"at Heidelberg, M\"onchhofstr. 12-14, D-69120 Heidelberg, Germany,
   \email{lemasle@uni-heidelberg.de} %1
        \and
          Astronomical Observatory, Odessa National University, Shevchenko Park, UA-65014 Odessa, Ukraine %2
        \and
            Dipartimento di Fisica, Universit\'a di Roma Tor Vergata, via della Ricerca Scientifica 1, I-00133 Rome, Italy %3
        \and
            INAF - Osservatorio Astronomico di Roma, via Frascati 33, Monte Porzio Catone, I-00078 Rome, Italy %4
        \and
            Agenzia Spaziale Italiana, Space Science Data Center, via del Politecnico snc, I-00133 Rome, Italy %5
        \and
            GEPI, Observatoire de Paris, CNRS, Universit\'e Paris Diderot, Place Jules Janssen, 92190, Meudon, France %6
        \and
            UPJV, Universit\'e de Picardie Jules Verne, 33 rue St. Leu, 80080, Amiens, France %7
        \and
            South African Astronomical Observatory, PO Box 9, 7935 Observatory, Cape Town, South Africa %8
        \and
            Southern African Large Telescope Foundation, PO Box 9, 7935 Observatory, Cape Town, South Africa %9
        \and
            Sternberg Astronomical Institute, Lomonosov Moscow State University, Universitetskij Pr. 13, Moscow 119992, Russia %10
             }

   \date{Received September 15, 1996; accepted March 16, 1997}

% \abstract{}{}{}{}{} 
% 5 {} token are mandatory
 
  \abstract
   {Mapping the Galactic spiral structure is a difficult task since the Sun is located in the Galactic plane and because of dust extinction. For these reasons, molecular masers in radio wavelengths have been used with great success to trace the Milky Way spiral arms. Recently, \textit{Gaia} parallaxes have helped in investigating the spiral structure in the Solar extended neighborhood.} 
   {In this paper, we propose to determine the location of the spiral arms using Cepheids since they are bright, young supergiants with accurate distances (they are the first ladder of the extragalactic distance scale). They can be observed at very large distances; therefore, we need to take the Galactic warp into account.} 
   {Thanks to updated mid-infrared photometry and to the most complete catalog of Galactic Cepheids, we derived the parameters of the warp using a robust regression method. Using a clustering algorithm, we identified groups of Cepheids after having corrected their Galactocentric distances from the (small) effects of the warp.}
   {We derived new parameters for the Galactic warp, and we show that the warp cannot be responsible for the increased dispersion of abundance gradients in the outer disk reported in previous studies. We show that Cepheids can be used to trace spiral arms, even at large distances from the Sun. The groups we identify are consistent with previous studies explicitly deriving the position of spiral arms using young tracers (masers, OB(A) stars) or mapping overdensities of upper main-sequence stars in the Solar neighborhood thanks to \textit{Gaia} data.
   }
   {}
  % conclusions heading (optional)

   \keywords{Stars: variables: Cepheids --
                Galaxy: disk --
                Galaxy: structure}

   \maketitle
%
%----------  INTRODUCTION  ----------

\section{Introduction}

\par Cepheids are massive or intermediate-mass pulsating variable stars; they are well-known as a calibrator of the extragalactic distance scale via their period-luminosity (PL) relations. Their ages range from a few tens to a few hundreds of megayears, which makes Cepheids excellent tracers of young stellar populations, for instance, in the Milky Way disk \citep[e.g.,][]{Lemasle2013,Ripepi2021}, in the Magellanic Clouds \citep[e.g.,][]{Lemasle2017,Romaniello2022}, and in nearby dwarf irregular galaxies \citep[e.g.,][]{Neeley2021}.

\par The Solar System is located close to the Milky Way plane \citep[z$_{\odot}$=20.8\footnote{We note in passing that z$_{\odot}$ values based on hydrogen radio emission are often smaller \citep[e.g., $\sim$4\,pc,][]{Blaauw1960} than those based on stellar tracers \citep[see also][]{bland2016}.}\,pc,][]{Bennett2019}, at  R$_{\odot}$=8.275\,kpc from the Galactic center \citep{GravityCollaboration2021}. Given the high extinction in the plane, mapping the Milky Way spiral structure is a difficult task. Maser sources associated with young massive stars in high-mass star-forming regions are among the most reliable tracers since they are very young and their parallaxes can be measured with radio-interferometry \citep{Reid2019}. Although radio-interferometric measurements are limited to a couple hundred sources, they present the strong advantage of being unaffected by extinction, and, therefore, of tracing the spiral structure at large distances from the Sun. On the basis of these measurements, \citet{Reid2019} found that the Milky Way spiral structure consists of four arms, plus the Local arm, which they consider to be an isolated segment. However, alternative models exist, for instance, a two-(major)-arm model by \citet{Drimmel2000}. \citet{Hou2014} demonstrate the difficulty in determining the number of spiral arms in the Milky Way. In this context, \textit{Gaia} constrained the location of several spiral arms in the fourth Galactic quadrant: within $\approx$5\,kpc from the Sun, where its parallaxes remain accurate enough, it provided distances for thousands of OB(A) \citep{Chen2019,Xu2021,Poggio2021,Zari2021} upper main-sequence stars or young open clusters \citep{Castro-Ginard2021,Hao2021,Monteiro2021}, all enabling various teams to trace the spiral arms in the extended Solar neighborhood. In this paper we take advantage of the accurate distances of classical Cepheids to investigate the spiral structure of the Milky Way. Since some of these Cepheids are very distant, we need to take the Galactic warp into account.

\par The paper is organized as follows: in Sect.~\ref{sect:cat}, we briefly explain how our catalog of classical Cepheids was gathered. In Sect.~\ref{sect:PL}, we take advantage of the catalog \citep[including \textit{Gaia} Early Data Release 3 (EDR3) data,][]{Gaia-edr3} to determine new period-luminosity and period-Wesenheit relations in the WISE bands, and to derive a homogeneous set of distances for the Cepheids. In Sect.~\ref{sect:warp}, we examine the properties of the Galactic warp. In Sect.~\ref{sect:spiral} we inspect the Milky Way spiral arms as traced by classical Cepheids, and their impact on abundance gradients is discussed in Sect.~\ref{sect:gradients}. Sect.~\ref{sect:conc} provides our summary and conclusions.

%----------  CATALOG  ----------

\section{The catalog}
\label{sect:cat}

\subsection{Input data: Variability catalogs}

\par We have built a comprehensive catalog of pulsating variable stars, gathering the data published by many photometric surveys dedicated to variability, or having at least some time-domain capabilities. They are listed below. 
\par One of our main sources of classical Cepheids is the Optical Gravitational Lensing Experiment (OGLE) survey, which provides a two-decade-long monitoring of the Magellanic Clouds \citep{Soszynski2015a,Soszynski2017a} and of the Milky Way bulge and disk \citep{Soszynski2017b,Soszynski2020,Udalski2018}. The data from the \textit{Gaia} satellite are an amazing all-sky tool to discover and monitor variable stars \citep{Clementini2019}. Due to the small number of observations covering their light curves, some variables were mis-classified in \textit{Gaia} DR2 \citep[see][for instance]{Lemasle2018}, which led \citet{Ripepi2019a} to reclassify the  \textit{Gaia} DR2 Galactic classical Cepheids, which we also added to our list of Galactic Cepheids. The All-Sky Automated Survey for Supernovae \citep[ASAS-SN,][]{Jayasinghe2018,Jayasinghe2019a,Jayasinghe2019b} surveys the entire sky down to V$\approx$18\,mag using a network of 24 small telescopes, making it particularly useful to follow (among others) the bright variables that are inaccessible to other surveys due to saturation. ASAS-SN draws from the All Sky Automated Survey \citep[ASAS,][]{Pojmanski1997,Pojmanski2002}, and we used additional data from the machine-learned ASAS Classification Catalog \citep[MACC,][]{Richards2012}. For bright Cepheids, we also used the list of classical Cepheids\footnote{\url{https://www.astrouw.edu.pl/ogle/ogle4/OCVS/allGalCep.listID}} provided by \citet{Skowron2019a}. The Zwicky Transient Facility \citep[ZTF,][]{Chen2020}, a recent time-domain survey, provided a large number of new targets, especially in the Northern Hemisphere, in general with an extremely good sampling of the period. Finally, the Wide-field Infrared Survey Explorer \citep[WISE,][]{Chen2018} discovered a large number of pulsating stars, and its observing window in the mid-infrared provides exquisite distances via period-luminosity or period-Wesenheit relations in a spectral domain where extinction is minimal. Since the classification of variable stars in the infrared is a challenging task (due to the similarity of the light curves of different classes of variable stars), we retained only those stars that could be identified as a classical Cepheid in at least one optical photometric survey.

\subsection{Merging the data and quality control}

\par Merging and quality control are important steps when collecting such an amount of data from various sources. We followed the procedure established for RR~Lyrae and Type~II Cepheids detailed in Lala et al., in prep. The catalog of classical Cepheids will be published in a forthcoming paper, which will also provide more details on the quality control process\footnote{and a comparison with the new list of classical Cepheids from the OGLE team \citep{Pietrukowicz2021}}. The main points are briefly listed below: 
\begin{itemize}[nosep]
    \item Stars from each individual survey have been robustly cross-matched against \textit{Gaia} EDR3, in order to recover their astrometry, and their photometry in the \textit{Gaia} bands (if available).
    Stringent quality cuts regarding large-scale systematics, binarity, or crowded regions using keywords such as \texttt{ruwe}, \texttt{ipd\_gof\_harmonic\_amplitude}, \texttt{astrometric\_excess\_noise}, etc were applied to \textit{Gaia} astrometric and photometric data following the recommendations of \citet{Fabricius2021,Lindegren2021a}, and \citet{Riello2021}.
    \item To avoid stars with spurious astrometric solutions, discarding stars with a fractional parallax uncertainty $>$\,0.15 (6.6$\sigma$) is common practice. However, we compared the parallax-based distances to those obtained from various period-luminosity or period-Wesenheit relations. The agreement is obviously better for stars with fractional uncertainties $<$\,0.15 (median difference $\sim$0.2\,kpc), but the agreement is still good for stars with fractional uncertainties between 3$\sigma$ and 6.6$\sigma$ (median difference $\sim$0.6\,kpc)\footnote{The extent of the agreement varies from star to star depending on the period-luminosity or period-Wesenheit relation used in the comparison}. 
    Therefore, we considered stars with a fractional parallax uncertainty $<$ 0.33. We note that this cut was not applied when deriving period-Wesenheit (PW) relations as it would bias the input sample. With such a criterion, we loose only a few tens nearby Cepheids that passed the previous quality cuts.
    \item The classification in different subclasses of pulsating variables and the exact value of the period has been taken from OGLE, and if not available, then from other surveys. If a star was observed in more than one survey, priority was given to the survey with the largest number of data points in the light curve. The classification and the value of the period are in general an excellent match between different surveys, and most of the confusion arises because the very nature of some of the surveys does not allow them to discriminate, for instance, anomalous Cepheids, or stars pulsating in several modes simultaneously.
    \item In addition, we checked for aliased periods (when the coverage of the light curve is inadequate, the recovered period may be a multiple of the true period). Similarly, candidate variables with a period matching exactly the terrestrial rotation period have been removed.
    \item For surveys observing in the same photometric bands, we checked for possible zero-point offsets between their photometry and found them to be negligible, with the exception of the MACC catalog.
\end{itemize}

%----------  PL/PW RELATIONS  ----------

\subsection{Period-Wesenheit relations and distances}
\label{sect:PL}
\begin{figure}[!ht]
  \includegraphics[width=\linewidth]{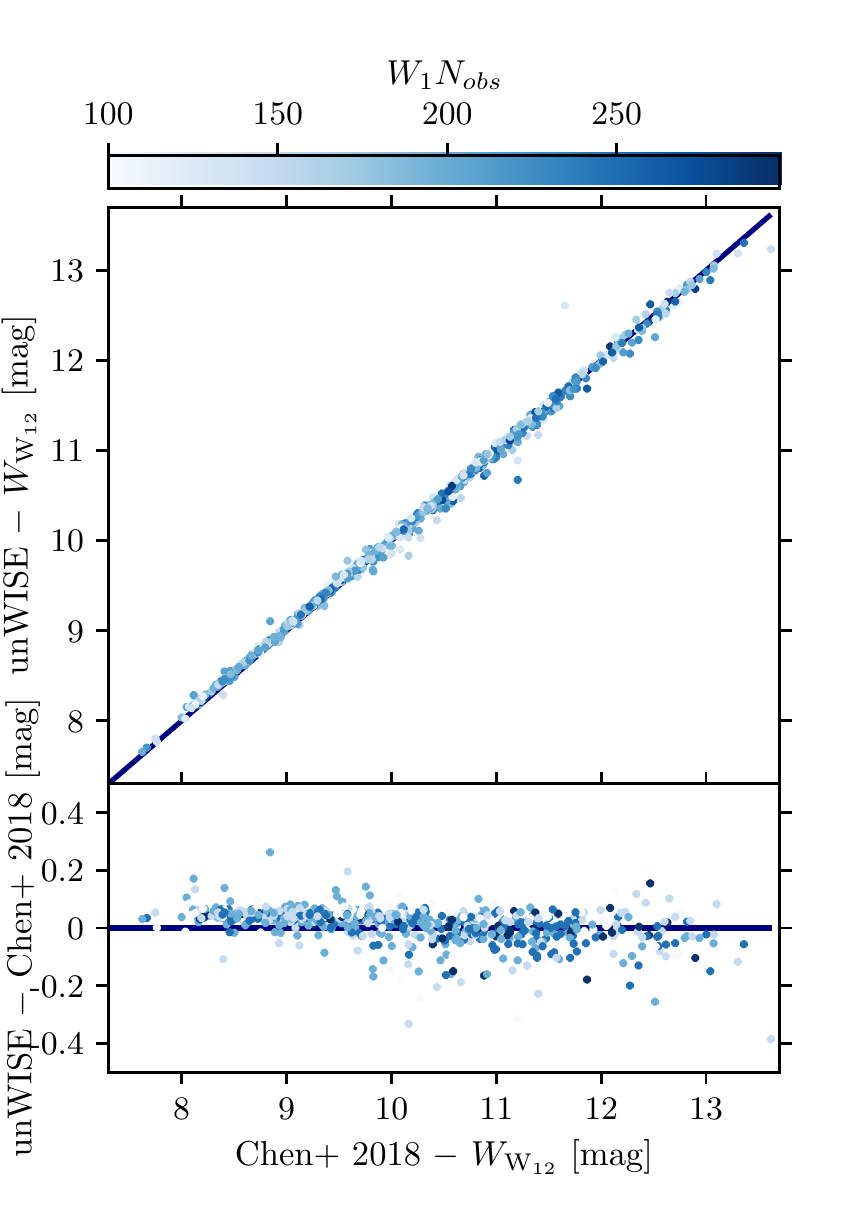}
  \caption{Comparison between the unWISE photometry (static coadds, 6 years of operations) and the WISE photometry in \citet{Chen2018} (light-curve fitting, 5 years of operations).}
  \label{fig:comp_WISE}
\end{figure}

We recovered unWISE photometry (W1, W2) for this catalog from IRSA\footnote{\url{https://irsa.ipac.caltech.edu/cgi-bin/Gator/nph-scan?utf8=\%E2\%9C\%93&mission=irsa&projshort=WISE}}  \citep{Meisner2021}. unWISE photometry consists of all-sky static coadds based on six years of WISE and NEOWISE operations, in contrast to the one-year ALLWISE data release \citep{Cutri2013}. Besides the longer observation baseline, unWISE photometry also holds the advantage of being more robust in crowded regions, on account of using crowdsource \citep{Schlafly2019} cataloging software. Fig.~\ref{fig:comp_WISE} compares unWISE photometry with that of the \citet{Chen2018} catalog. The latter presented an all-sky variable star catalog based on five years of WISE and NEOWISE data and their magnitudes were determined by Fourier-fitting the light curves. The greater number of measurements, coupled with the fact that the pulsation amplitude of classical Cepheids is roughly 0.2\,mag \citep[e.g.,][]{Chen2018} in mid-IR wavelenghts, results in unWISE photometry agreeing excellently with Fourier-fitted mean magnitudes. We obtained W1, W2 unWISE photometry for 3260 Cepheids (after taking all the processing flags\footnote{\url{https://catalog.unwise.me/files/unwise_bitmask_writeup-03Dec2018.pdf}} into account).

To determine distances using unWISE photometry, we computed new period-Wesenheit relations in WISE bands. We created a catalog of LMC classical Cepheids, similar to the one described here for the Milky Way. The apparent Wesenheit $W_{\mathrm{W12}}$ was calculated as:
\begin{equation}
W_{\mathrm{W_{12}}} = W_2 - R_{\mathrm{W_2, W_1}} * (W_1 - W_2).    
\end{equation}
The multiplicative constant $R_{\mathrm{W_2, W_1}}$ is the total-to-selective extinction ratio:
\begin{equation}
R_{\mathrm{W_2, W_1}} = \frac{A_\mathrm{W_2}}{E(W_1-W_2)},
\end{equation}
with R$_{\mathrm{W_2, W_1}}$ = 2.0 \citep[][]{WangChen2019}. Absolute Wesenheits were calculated using the LMC distance modulus (18.477\,mag) from \citet{Pietrzynski2019}. We used \texttt{pymc3} \citep{Salvatier2016} to perform a Bayesian robust regression (as described in Sect.~\ref{sect:robust}) on the following model:

\begin{equation}
W_{\mathrm{abs}} \sim \mathcal{T}(\alpha + \beta \times \mathrm{log}_{\mathrm{10}}\left\{\frac{\mathrm{Period}}{1 \mathrm{d}}\right\}, \, \sigma^{2}, \, \nu),
\end{equation}

\noindent where $\alpha$ and $\beta$, the intercept and the slope of the model, are assumed to follow a normal distribution, while $\sigma$, the intrinsic scatter, is assumed to follow a half-normal distribution and $\nu$, the normality parameter (degrees of freedom), is assumed to follow a Gamma distribution \citep{Juarez2010}. $ W_{\mathrm{abs}}$ is the absolute Wesenheit and we assume that the likelihood of the model follows a Student's T-distribution. The model parameters (and their uncertainties) for the PW relations in WISE bands are given in Table~\ref{tab:params_dist_regression}. The covariance matrix is provided in Table~\ref{tab:cov_matrix_PW}. From these relations, we obtained distances precise up to 3\% (5\%) in the low- (high-) extinction regions. The Wesenheit pseudo-magnitudes are unaffected by the extinction toward individual stars (or their uncertainties) by construction \citep{Madore1982}. Their dependence on reddening only lies in the accuracy and potential nonuniversality of the R$_{\mathrm{W_2, W_1}}$ ratio.

\begin{table}[!ht]
\centering
\caption{Period-Wesenheit relations in WISE bands for fundamental mode (DCEP\_F) and first-overtone (DCEP\_10) classical Cepheids located in the LMC.}
\label{tab:params_dist_regression}
\resizebox{0.495\textwidth}{!}{
\begin{tabular}{lcccc}
\hline\hline
Type & $n$ & $\alpha$ & $\beta$ & $\sigma$ \\
\hline
DCEP\_F & 2326 & -2.436 $\pm$ 0.013 & -3.196 $\pm$ 0.019 & 0.149 $\pm$ 0.005 \\
DCEP\_1O & 1591 & -2.936 $\pm$ 0.015 & -3.342 $\pm$ 0.037 & 0.181 $\pm$ 0.008\\
\hline
\end{tabular}
}
\tablefoot{$n$ denotes the total number of stars used to compute the law. The mean and standard deviations of posterior distributions of the model parameters are presented in the last three columns}
\end{table}

\begin{table}[ht]
\centering
\scriptsize
\caption{Covariance matrix for the Bayesian robust regression of the period-Wesenheit relations.}
\label{tab:cov_matrix_PW}
\begin{tabular}{rrrrr}
\hline\hline
       & \multicolumn{1}{c}{$\alpha$} & \multicolumn{1}{c}{$\beta$} & \multicolumn{1}{c}{$\sigma$} & \multicolumn{1}{c}{$\nu$} \\
\hline
$\alpha$ &  3.24967949e-04 & -4.38328130e-04 &  1.46899174e-03 & -6.11806477e-05 \\
 $\beta$ & -4.38328130e-04 &  6.77959533e-04 & -2.16090433e-03 &  2.92371896e-05 \\
$\sigma$ &  1.46899174e-03 & -2.16090433e-03 &  1.52551182e+00 &  1.10784521e-03 \\
   $\nu$ & -6.11806477e-05 &  2.92371896e-05 &  1.10784521e-03 &  2.26928615e-03 \\
\hline
$\alpha$ &  3.08465117e-04 & -6.95175096e-04 & -2.30024957e-04 & -1.07183562e-04 \\
 $\beta$ & -6.95175096e-04 &  2.01753200e-03 &  2.93887199e-03 &  1.97729713e-04 \\
$\sigma$ & -2.30024957e-04 &  2.93887199e-03 &  1.38204481e+00 &  2.73979159e-03 \\
   $\nu$ & -1.07183562e-04 &  1.97729713e-04 &  2.73979159e-03 &  2.53117985e-03 \\ 
\hline
\end{tabular}
\tablefoot{ $\alpha$ and $\beta$ are the zero-point and the slope of the relations, $\sigma$ their standard deviation and $\nu$ their normality parameter. The upper panel is for fundamental mode classical Cepheids and the lower panel for first-overtone classical Cepheids.}
\end{table}

\par In what follows, we used only Cepheids pulsating in the fundamental (F) or the first overtone (1O) mode: they are, by far, the most numerous, and Cepheids in other subclasses have slightly less accurate distances since their period-Wesenheit relations are calibrated with a smaller number of stars. We used distances determined using the best WISE data, as described above, for 2098 Cepheids for which such photometry was available. If not (586 Cepheids), we used as distance the inverse of the \textit{Gaia} EDR3 parallax, provided this distance is less than 5\,kpc. Uncertainties on the astrometric or photometric distances have been propagated throughout the paper. This restricted catalog contains 2684  Cepheids (F,1O), while the catalogs of \citet{Chen2019} and \citet{Skowron2019b} contained 1339 and 2390 Cepheids, respectively.

%----------  WARP  ----------

\section{The Galactic warp as traced by classical Cepheids}
\label{sect:warp}

\subsection{The Galactic warp}

\par The Galactic warp \citep[][]{Kerr1957,Oort1958} is a large-scale distortion of the Milky Way disk. It is caused by a torque exerted on the disk, whose origin has been suggested to result either from a misalignment between the rotation axis of the disk and of the halo \citep[e.g.,][]{Sparke1988,debattista1999}, or from the inner disk \citep[e.g.,][]{Chen2019}, or from material accreted to the halo \citep[e.g.,][]{Ostriker1989,Jiang1999}, or from tidal perturbations associated with nearby Milky Way satellites such as Sagittarius \citep[e.g.,][]{ibata1998,Laporte2019} and the Magellanic Clouds \citep[e.g.,][]{Weinberg2006,Garavito2019}. 

\par Different tracers have been used to map the warp, from neutral hydrogen \citep[e.g.,][]{Henderson1982} to molecular clouds \citep[e.g.,][]{Wouterloot1990} and star counts \citep[e.g.,][]{Reyle2009,Amores2017}. Individual stellar classes, namely OB stars \citep[e.g.,][]{Miyamoto1988,Reed1996,Yu2021}, RGB or red clump stars \citep[e.g.,][]{Lopez-Corredoira2002,Momany2006,Wang2020}, and even pulsars \citep[e.g.,][]{Yusifov2004} have also been used. It has been shown in the Milky Way \citep[e.g.,][]{Chen2019,Romero-Gomez2019} and in external galaxies \citep[e.g.,][]{Radburn-Smith2014} that only the youngest stellar objects accurately map the Galactic warp. The mismatch between the warp as traced by hydrogen and young stellar objects, and the one traced by older stars, suggests a large value for the warp's precession, leading, for instance, \citet{Poggio2020} and \citet{Cheng2020} to favor the tidal perturbation scenario. On the other hand, \citet{Chrobakova2021} found no evidence of a warp precession.

\par Cepheids are an ideal tracer for studying the present-day warp because they are easily distinguished from other types of stars, and because their distances can be derived individually with great accuracy, even at large distances ($>$5~kpc) where \textit{Gaia} parallaxes become uninformative. \citet{Chen2019,Skowron2019a,Skowron2019b} have used distances derived from mid-infrared (Spitzer: \citet{Benjamin2003,Churchwell2009}, WISE: \citet{Chen2018}) or near-infrared photometry (2MASS: \citet{Skrutskie2006}). \citet{Dekany2019} traced the warp in highly reddened regions covered by the VVV survey. All these studies enabled the tracing of the Galactic warp and they confirmed that the Cepheids' warp follows the \ion{H}{i} warp. Moreover, \citet{Skowron2019b} showed that the northern part of the warp is very prominent, with an amplitude 10\% larger than that of the southern warp.

\par These studies differ regarding the analytical formula adopted for the warp, the input catalog of Cepheids (different surveys have different completeness and contamination levels), and the photometry used to compute the Cepheids' distances. With respect to those studies, we benefit from updated WISE data, and our sample is larger (by several hundreds of stars) without sacrificing purity. We adopt the definition of the warp proposed by \citet{Skowron2019b}: the Milky Way warp starts at a given radius $r_{0}$ and its shape follows the equation:
\begin{equation}
\label{eq:warp}
\resizebox{.93\linewidth}{!}{$
z(r,\Theta)=
    \begin{cases}
        z_{0} & r < r_{0} \\
        z_{0}+(r-r_{0})^{2}\times[z_{1}\,sin(\Theta-\Theta_{1})+z_{2}\,sin(2\,(\Theta-\Theta_{2}))] & r \geq r_{0}
\end{cases}$}
%\text{for}~
\end{equation}
where $z$ is the vertical distance from the Galactic plane, $r$ is the distance from the Galactic center, and $\Theta$ is the Galactocentric azimuth. $\Theta$=0$^\circ$ points in the "Sun to Galactic center" direction, while $\Theta$=180$^\circ$ points toward the Galactic anticenter. $\Theta$ increases counterclockwise if the Galaxy is seen from above. We adjusted the radial, vertical, and angular parameters $r_0$, $z_0$, $z_{1}$, $z_{2}$, $\Theta_{1}$, $\Theta_{2}$, using a Bayesian robust regression method.

\subsection{Tracing the warp}
\label{sect:robust}

\par We estimated the parameters of the warp model using Bayesian robust regression. As mentioned above, we assume the warp formula by \citet{Skowron2019b} given in Eq.~\ref{eq:warp} for the likelihood of our model. We assume a Student's t-distribution for the likelihood of our model, as it is much less sensitive to outliers and, therefore, provides more robust estimates of the parameters than those from a normal distribution (in presence of outliers, other approaches often shift the mean toward the outliers and increase the standard deviation, while the Student's t-distribution decreases the weight of the outliers). We use the Hamiltonian MCMC sampler \citep{Betancourt2017} of \texttt{pymc3} to sample the posterior distribution. 
Uncertainties on (r,$\Theta$,z) and their covariances have been propagated from the uncertainties and covariances on right ascension, declination, and distances using Jacobian matrices \citep[][]{ESA1997,Gala} to perform coordinates transformations (the uncertainties on the position of the Sun relative to the Galactic center have been ignored).

\par We run the analysis in two steps: first we assume a normal distribution for the priors, adopting for the mean value and the standard deviation $r_0$=5$\pm$2\,kpc, $z_0, z_1, z_2$=0.5$\pm$1\,kpc,  $\Theta_{1}, \Theta_{2}$=$\pi\pm\pi$\,rad. From the posterior distributions of this first stage, we use the mean values and 10$\times$ the standard deviations as the input (normal) priors for the second stage. In both steps, we run 4 chains and use the first 10000 samples of each chain to tune the multidimensional posterior and accept the next 5000 draws as our posterior distribution (we checked that the auto-correlation is low for each individual chain). From the posterior distributions obtained in the second step, we adopt the mean values and the standard deviations as the parameters of the warp and their uncertainties, respectively. They are listed in Table~\ref{tab:params_bayesian}. The posterior distributions of the parameters, as well as the standard deviation of the regression model $\sigma$ are is shown in Fig.\ref{fig:corner_plot_warp}, which was drawn using the \texttt{ArviZ} package \citep{arviz_2019}. The full covariance matrix is provided in Table~\ref{tab:cov_matrix_warp}.

\begin{table}[!hb]
\centering
\caption{Parameters of the Galactic warp as derived using a robust regression method.}
\label{tab:params_bayesian}
\begin{tabular}{rrr}
\hline\hline
{} & \multicolumn{1}{c}{Mean} & \multicolumn{1}{c}{$\sigma$} \\
\hline
%$r_{0}$ (kpc) &   4.863 &    0.314 \\
%$z_{0}$ (kpc) &   0.013 &    0.005 \\
%$z_{1}$ (kpc) &   0.009 &    0.001 \\
%$z_{2}$ (kpc) &   0.001 &    0.000 \\
$r_{0}$ (kpc) &   4.8626 &    0.3136 \\
$z_{0}$  (pc) &     13.0 &       4.9 \\
$z_{1}$  (pc) &      8.9 &       0.6 \\
$z_{2}$  (pc) &      1.4 &       0.3 \\
$\theta_{1}$ (deg) & -13.48 &    1.92 \\
$\theta_{2}$ (deg) & -26.27 &    5.60 \\
$\sigma$ ~~~~~~~~~ &   0.052 &    0.003 \\
\hline
\end{tabular}
\end{table}

\par
\begin{figure*}[!ht]
  \includegraphics[width=\linewidth]{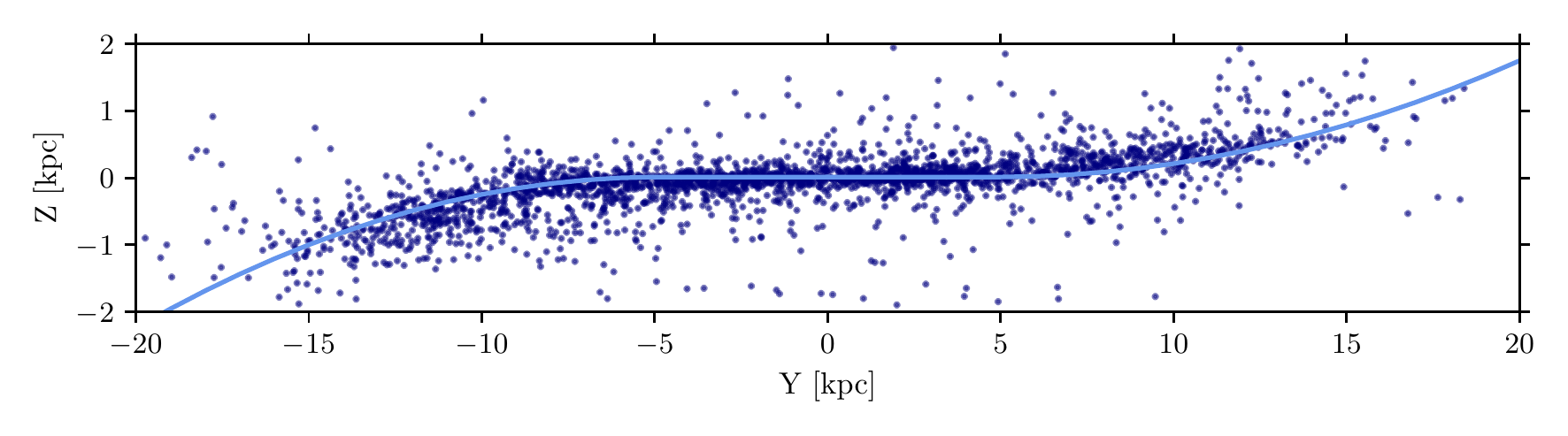}
  \caption{Milky Way warped disk computed with the parameters obtained from the robust regression method (light blue) for X=0\,kpc. Individual Cepheids are over-plotted in dark blue, with Galactocentric distances computed using WISE mid-infrared photometry and PW relations. See also Fig.~\ref{fig:warp_sectors} in the appendix.}
  \label{fig:warp}
\end{figure*}

 The model of the warp computed with these parameters is displayed in Fig.~\ref{fig:warp}, where the scale of the vertical axis is strongly enhanced. This figure, as well as Fig.~\ref{fig:warp_sectors} in the Appendix, shows that our model reproduces closely the vertical distribution of the Galactic Cepheids. The warp is more pronounced than the one provided by \citet{Chen2019}, and this was already noted by \citet{Skowron2019b}. The reason is simply that both our study and the one by \citet{Skowron2019b} rely on a larger number of Cepheids covering the four Galactic quadrants (although the sample is clearly incomplete in Q1 and Q4), while the Cepheids in \citet{Chen2019} mostly belong to Q2 and Q3.

\par The onset radius $r_{0}$ of the warp in our study is in fairly good agreement with the value reported by \citet{Skowron2019b} (4.86$\pm$0.31\,kpc vs. 4.23$\pm$0.12\,kpc). 
We note that if the formal value of the onset radius of the warp is small, the influence of the warp on the vertical position of Cepheids starts to be noticeable only at roughly the Solar radius.

\par The vertical parameters $z_{1}$ and $z_{2}$ are very similar in both studies: $\sim$7 and  $\sim$1\,pc in our case vs $\sim$8 and $\sim$2\,pc for \citet{Skowron2019b}. Our value for $z_{0}$  ($\sim$26\,pc) is smaller than the $\sim$44\,pc reported by \citet{Skowron2019b}, but in good agreement with recent literature values, for instance, the G+early K stars in the nearby sample of \citet{GaiaSmart2021}. 

\par Due to differences in their definition, Galactocentric azimuths are shifted by 180$^\circ$ between our work and \citet{Skowron2019b}. 
Our value of $\mathrm{\Theta_{1} = -13.48^\circ}$ must then be compared to their $\mathrm{\Theta_{1} = 158.3-180=-21.7^\circ}$, while the values of $\mathrm{\Theta_{2}, -26.27^\circ~\text{and} -13.6^\circ}$, respectively can be compared directly given the factor 2 in the second sine term of Eq.~\ref{eq:warp}. The angular values are not exact matches, they remain however close to each other and lead to a very similar description of the Galactic warp.

\par It is not a surprise that our model resembles the warp model by \citet{Skowron2019b}: we adopted their analytical relation, and even if we added a few hundreds of stars, the overall coverage of the disk is similar. However, the technique to derive the warp parameters is completely independent\footnote{To derive the warp parameters, \citet{Skowron2019a,Skowron2019b} simply mention that they minimize the sum of squares of orthogonal distances between individual Cepheids and the model, with the squared distances modulated by an exponential term penalizing outliers.}. We believe that the small differences originate from this somewhat larger stellar sample combined to the updated WISE photometry.
Moreover, uncertainties on coordinates and distances are included in the fitting procedure in our study. Although not formally compatible with those of \citet{Skowron2019b}, the uncertainties on the warp parameters are extremely small. Considering the spatial extent of the disk, they do not impact the determination of the Cepheids' Galactocentric distance.

Our results are also in excellent agreement with studies of the warp using \ion{H}{I} data. For instance, \citet{Nakanishi2003} found that the warping in \ion{H}{I} is the strongest for $\theta$=+80$^\circ$ and $\theta$=+260$^\circ$. We find similar angular values (see Fig.~\ref{fig:warp_sectors}). They report that the warping starts at  R$_{G}$=10-12\,kpc and reaches $\sim$1.5\,kpc at R$_{G}$=16\,kpc ($\theta$=+80$^\circ$) and $\sim$-1\,kpc at R$_{G}$=16\,kpc ($\theta$=+260$^\circ$). \citet{Levine2006a} focused on the \ion{H}{I} outer disk, well beyond the stellar disk. They found the maximum warping at $\theta$=+90$^\circ$ and $\theta$=270$^\circ$. The height of the warp reaches $\sim$4\,kpc at R$_{G}$=22\,kpc and $\sim$5.5\,kpc at R$_{G}$=28\,kpc. At R${_G}$=16\,kpc, they report a maximum height of $\sim$1.3\,kpc and a maximum depth of $\sim$-0.8\,kpc. These values, later confirmed by \citet{Kalberla2007}, are slightly lower than those provided earlier by \citet{Nakanishi2003} and in better agreement with our own findings. \citet{Levine2006a} found that the \ion{H}{I} warp can be approximated with a superposition of three vertical harmonics of the disk. Interestingly, these modes grow linearly in the outer disk. The mode m=1, which dominates the warp for the range of Galactocentric distances covered by our Cepheids' sample, is linear between R$_{G}$=10\,kpc and R$_{G}$=25\,kpc with a slope of 0.197\,kpc\,kpc$^{-1}$, which leads to a height of $\simeq$1\,kpc at R${_G}$=15\,kpc. The other two modes do not influence the warp below R${_G}$=15\,kpc, and they grow linearly with similar slopes until R$_{G}$=22\,kpc.

\begin{figure}[!ht]
  \includegraphics[width=\linewidth]{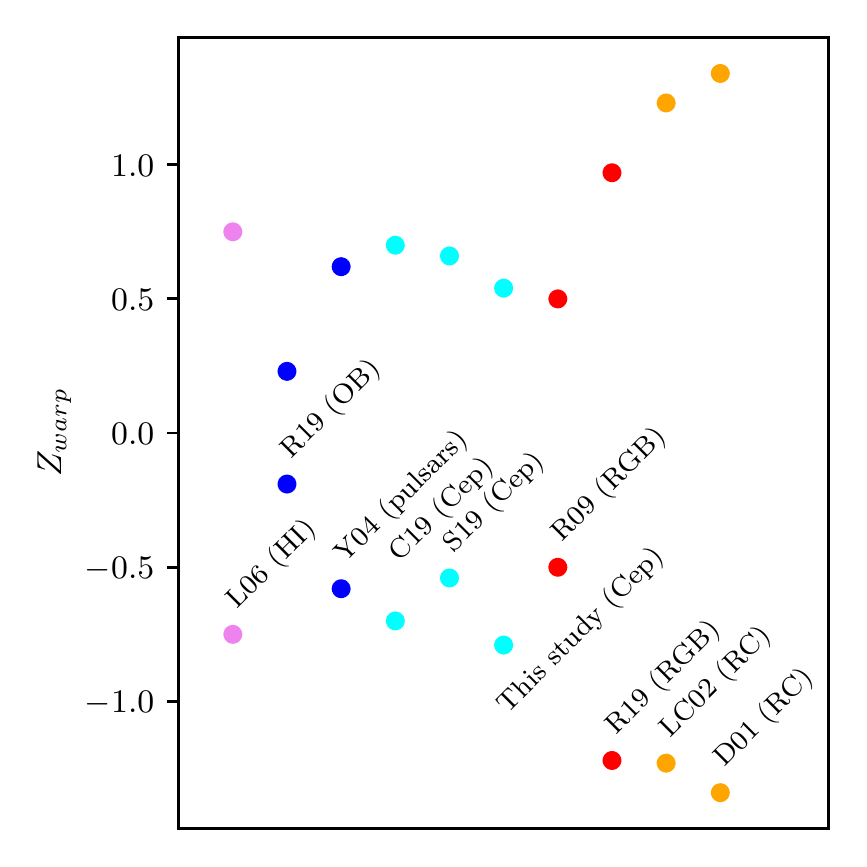}
  \caption{Values of the warp altitude above and below the Galactic plane at the Galactocentric radius R=14\,kpc for different tracers ordered by approximate age. The values have been taken from Table~1 in \citet{Romero-Gomez2019} or computed by us. The tracers used to investigate the warp cover neutral hydrogen \ion{H}{i} \citep[][L06]{Levine2006b}, OB stars \citep[][R19]{Romero-Gomez2019}, pulsars \citep[][Y04]{Yusifov2004}, Cepheids (\citet[][C19]{Chen2019}, \citet[][S19]{Skowron2019b}, this study), RGB (\citet[][R09]{Reyle2009}, \citet[][R19]{Romero-Gomez2019}), and red clump (RC) stars (\citet[][D01]{Drimmel2001}, \citet[][LC02]{Lopez-Corredoira2002}).}
  \label{fig:warp_height}
\end{figure}

A comparison of the warp altitudes above and below the Galactic planes reached by various tracers at a Galactocentric distance of R=14\,kpc is shown in Fig.~\ref{fig:warp_height}. It indicates that the warp becomes more pronounced for older tracers, as already mentioned, for instance, by \citet[][]{Romero-Gomez2019}.
 
\subsection{Unwarping the Milky Way}
\label{sect:unwarping}

\par In abundance gradient studies, the abundance of a given element is plotted against the radial Galactocentric distance of the stars composing the sample, or against their distance to the Galactic plane. However, in regions where the disk is strongly warped, the Galactocentric distances of stars end up being shorter than if the star was located in a nonwarped Milky Way disk. Because of the scarcity of spectroscopic data, very few studies have analyzed the spatial distribution of abundances \citep[see][for instance]{Kovtyukh2022}. Instead, it has been customary to collate Cepheids located at different Galactocentric azimuths in a unique 2D plane, where [Fe/H] is displayed as a function of the Galactocentric distance of the Cepheids in the sample. This necessary shortcoming implies ignoring the warping of the disk. It might be (at least partially) responsible for the increased dispersion of abundances in the outer disk since it brings together Cepheids located in warped and nonwarped regions. 
\par In order to investigate this issue, we have compared the length of a bow between the Galactic Center and a Cepheid located at the Galactocentric distance d$_{\mathrm{GC}}$ with d$_{\mathrm{GC}}$ itself. The details of the calculation are given in Appendix \ref{App:unwarp}. As can be seen in Fig.~\ref{fig:Unwarp_figure}, the difference between Galactocentric distances computed on a flat and on a warped disk are negligible below 10\,kpc. They can reach 100\,pc at R$_{G}$=18\,kpc, but only for the stars located in regions where the warp is pronounced, while other Cepheids are barely affected. The differences become larger only in the outer regions of the disk, where only a small number of Cepheids have been reported until now. They remain, however, too small to explain the larger dispersion around the mean metallicity gradient reported, for instance, by \citet[][]{Genovali2014} in the outer disk.

\begin{figure}[!ht]
\includegraphics[width=\linewidth]{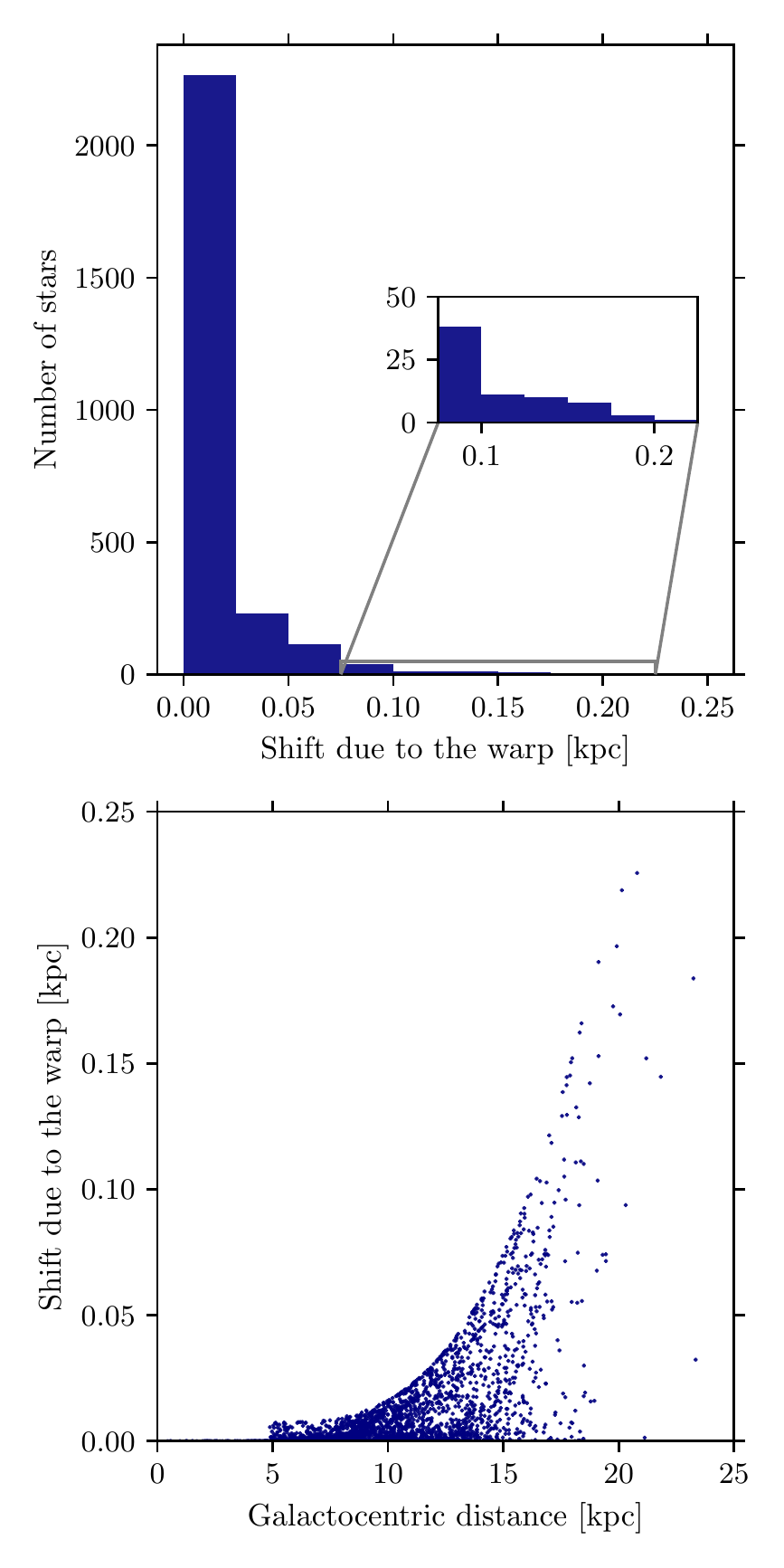}
\caption{Impact of the warp on Galactocentric distances. {\it Top panel:} Distribution of the difference between Galactocentric distances computed on a flat and on a warped disk for our entire sample of Cepheids. {\it Bottom panel:} Difference between Galactocentric distances computed on a flat and on a warped disk, as a function of the Galactocentric distance.}
\label{fig:Unwarp_figure}
\end{figure}  

%----------  Spiral arms  ----------

\section{Tracing the spiral arms with classical Cepheids}
\label{sect:spiral}

\par Since they provide accurate distances, there have been many investigations of the Galactic structure using Cepheids, often with the primary goal to derive the Galactocentric distance of the Sun or to determine the Milky Way rotation curve \citep[e.g.,][and references therein]{Caldwell1987,Pont1997,Metzger1998,Mroz2019}. Regarding the spiral arms, \citet{Dambis2015} matched their Cepheid data to a four-armed pattern with a pitch angle of 9.5$\pm$0.1$^{\circ}$.
Using Cepheids in the far side of the disk, \citet{Minniti2021} favor instead a two-arms model expanding into four arms for $R_{G}\gtrapprox$~5--6\,kpc. However, the spiral structure has mostly been traced by younger tracers \citep[for instance, \ion{H}{ii} regions,][]{Georgelin1976}, although their distances were in general less accurate because they rely on kinematical models. Indeed, in a traditional textbook picture of a spiral arm, a shock wave concentrates material in the so-called dust lane. Toward the outer disk, one then encounters masers associated to protostars, followed by \ion{H}{ii} regions, and further on by stars having reached (e.g., OB stars) or evolved off the main sequence \citep[e.g.,][]{Roberts1969,Vallee2020}.
\subsection{The age question and the choice of spiral arms tracers}
\label{sect:age_discussion}

\par It was quickly suggested \citep{Fernie1958,Kraft1963} that the brighter, longer-period Cepheids match the spiral arms as traced by atomic hydrogen better than their fainter, shorter-period counterparts. It was also correctly conjectured that such Cepheids are younger and, therefore, had less time to drift away from their birthplace. Indeed the age of a Cepheid is inversely correlated with its period via period-age relations \citep[see][and references therein]{Efremov1978}. Two ways can be envisioned to overcome the issue of tracers of the spiral arms having evolved off them: either selecting truly young tracers (\ion{H}{ii} regions, O stars), or selecting only the youngest objects for tracers spanning a larger age range.

\par For instance, \citet{Castro-Ginard2021} restricted their sample of open clusters to those younger than 80\,Myr, while \citet{Hao2021} used 100\,Myr. Selecting young stellar groups within 3\,kpc from the Sun, 
\citet{Kounkel2020} report that the separation between spiral arms remains visible up to 63\,Myr (log(age)=7.8). Their scenario favors transient arms, and indicates that the Sagittarius arm has moved toward the Galactic center by 0.5\,kpc in the last 100\,Myr. Using a small number of Cepheids in the Solar neighborhood, also split into a younger and an older group, \citet{Veselova2020} found 7 and 8 spiral arm segments, respectively. For a given segment, they report that the parameters retrieved for the young and the old objects differ, especially in the case of the Sagittarius and the Perseus arms \citep[see also][]{Bobylev2021}.

\par In this context it is worth mentioning that Cepheids' ages are not extremely accurate since they can vary by a factor up to 2 depending on whether stellar rotation is included \citep{Anderson2014a} or not \citep{Bono2005} in the evolutionary models \citep[see also the recent paper by][with no rotation]{DeSomma2020b}. However, their ranking by age is very reliable since their period, which can be measured with great accuracy, is the driving parameter via period-age relations. In an attempt to constrain these relations using Cepheids in open clusters, \citet{Medina2021} noted that ages of young open clusters potentially hosting Cepheids (and, therefore, younger than $\approx$300\,Myr) are quite inaccurate given the paucity or even the absence of evolved stars and the stochastic sampling of their initial mass function (IMF). Such uncertainties affect not only the absolute ages of young clusters but also their ranking by age.

\par Both \citet{Poggio2021} and \citet{Zari2021} found overdensities of upper main-sequence stars in \textit{Gaia} data, which could be associated with the Sagittarius-Carina and the Scutum-Centaurus arms. \citet{Poggio2021} found no obvious match between their overdensities and Cepheids with age $<$100\,Myr, while in contrast the agreement was good with open clusters of similar ages. They noted, however, that the young Cepheids ($<$100\,Myr) overlap quite well with the spiral structure proposed by \citet{Taylor1993} or the Perseus arm characteristics proposed by \citet{Levine2006b}. \citet{GaiaDrimmel2022} reached the same conclusions, we note in passing that they accepted Cepheids up to 200\,Myr old in their sample. Similarly, \citet{Majaess2009} indicate that Cepheids younger than 80\,Myr in their sample are good tracers of the spiral structure. Although the same level of detail cannot be reached at large distances, it is worth mentioning that in M31, the sample of classical Cepheids of \citet{Kodric2018a} closely follows the position of the ring structures rich in dust and star-forming regions. Finally, \citet{Minchev2013,Minchev2014} coupled their chemo-dynamical model to high-resolution simulations tailored to the Milky Way in a cosmological context. They concluded that the oldest stars are the most affected by stellar radial migration, while the young stars are found near their birth radii. Recent studies \citep[e.g.,][and references therein]{Frankel2020,Lian2022} confirmed that radial migration is inefficient over short time-scales\footnote{\citet{Lian2022} report, for instance, average migration distances of 0.5-1.6\,kpc after 2~Gyr and 1.0-1.8\,kpc after 3~Gyr (roughly 20 to 150 times more than the ages of Cepheids considered here)}.

\par
\begin{figure}[!ht]
    \centering
    \includegraphics[width=\linewidth]{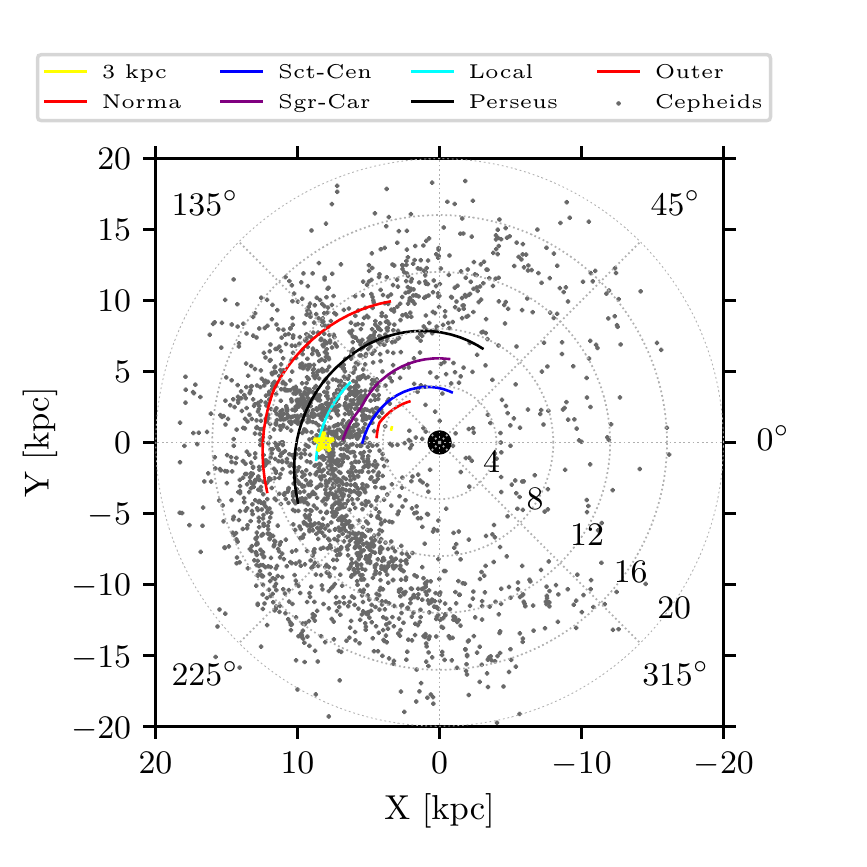}
    \caption{Classical Cepheids (black dots) in the Galactic plane. The Galactocentric distances are derived from mid-infrared photometry to minimize the effect of reddening. The spiral arms delineated by \citet{Reid2019} are over-plotted using the same color-coding as in the original paper. Two segments are plotted in red as they presumably belong to the same Norma-Outer ring. Concentric circles are shown every 4\,kpc to guide the eye. The Galactic center (black filled circle) is at (0,0) and the Sun (yellow star) at (8.275,0). 
    }
    \label{fig:spiral_Reid19}
\end{figure}

\begin{figure}[!ht]
    \centering
    \includegraphics[width=\linewidth]{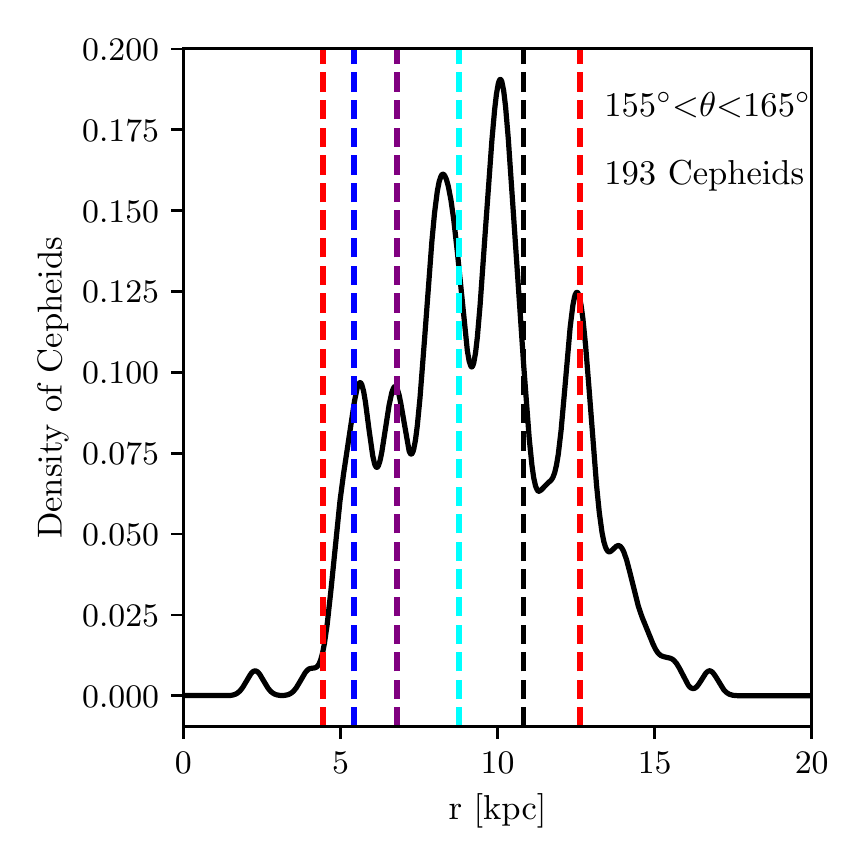}
    \caption{
    Kernel density estimation (with a kernel bandwidth of 0.1) of the radial distribution of Cepheids located in a Galactocentric angular sector around 160$^\circ$. The location of the spiral arms of \citet{Reid2019} in this sector are shown using the same color-coding as in the original paper. The angular sector around 160$^\circ$ intercepts the spiral arms in a region where the completeness of the data is not hindered by the two shadow cones visible in Fig.~\ref{fig:spiral_Reid19} which hamper the detection of Cepheids beyond nearby regions with strong extinction.}
    \label{fig:Cepheid_density}
\end{figure}

 Notwithstanding, over-plotting the spiral arms delineated by \citet{Reid2019} on top of the entire sample of Cepheids without any age restriction, as shown in Fig.~\ref{fig:spiral_Reid19}, indicates that the Cepheids' overdensities match the spiral arms well. Inter-arm regions have lower densities of Cepheids, as can be seen in the radial distribution of Cepheids located in a Galactocentric angular sector around 160$^\circ$ displayed in Fig.~\ref{fig:Cepheid_density}.

\par In what follows, we use the analytical period-age relations provided by \citet{Bono2005} to derive the Cepheids' ages, and we restrict our sample to stars pulsating in the fundamental or the first-overtone mode for which such relations are available\footnote{Second-overtone and multimode Cepheids are quite uncommon in the Milky Way, see \citet{Bono2002d,Smolec2010,Lemasle2018}}.

\subsection{Locating groups of Cepheids}

\par To identify the spiral arms, we used \texttt{t-SNE}  \citep[t-distributed Stochastic Neighbor Embedding,][]{vanDerMaaten2008}, a nonlinear dimensionality reduction technique. Although \texttt{t-SNE} is often used to visualize high-dimensional data in a lower-dimensional space, we only used as input the coordinates ($\theta, ln r$) of the Cepheids in our dataset, where $r$ has been corrected from the effects of the warp (see Sect.~\ref{sect:unwarping}). Since the algorithm uses a Student's t-distribution to compute the similarity between two data points in the \texttt{t-SNE} output space, it performs very well in keeping similar input data points close together in the output space, even if they come from crowded regions. The downside is that \texttt{t-SNE} performs poorly when data are sparse. After the data were standardized, \texttt{t-SNE} was initialized using a principal component analysis and run for 6000 iterations in a 2D space. The perplexity (the effective number of neighbors considered by \texttt{t-SNE} for any given data point) was set to 90. For our dataset, the topology of the outcome in the \texttt{t-SNE} space is robust to the choice of the perplexity value, as well as to the value of the early exaggeration (set to 5), which ensures that tight clusters in the data will not overlap in the \texttt{t-SNE} space. Individual groups are then identified using the clustering algorithm \texttt{HDBSCAN} (see details below).

\par To ascertain that our \texttt{t-SNE+HDBSCAN} algorithm is working, we have run several tests where the mock spiral structure is based on the \citet{Reid2019} model. They are described in detail in Appendix~\ref{App:test_algo}. Tests show that the algorithm recovers the mock spiral arms very well, even in the presence of large amounts of "inter-arm" Cepheids, that can be considered as noise. A fraction of these "inter-arm" Cepheids is then included in the nearest arm, but this impacts the recovered location of the spiral arm only marginally. We note that the algorithm is sensitive to small gaps (regions without stars) in individual spiral arms. A given spiral arm may then be split in several segments limited by those gaps. This is more likely to occur when two spiral arms are very close to each other. In such a case, it might even happen that two segments from two different arms are wrongly joined within the same group (and the recovered arm location wrongly falls at a median distance between the two segments).

\par
\begin{figure*}[!ht]
    \centering
    \includegraphics[width=\linewidth]{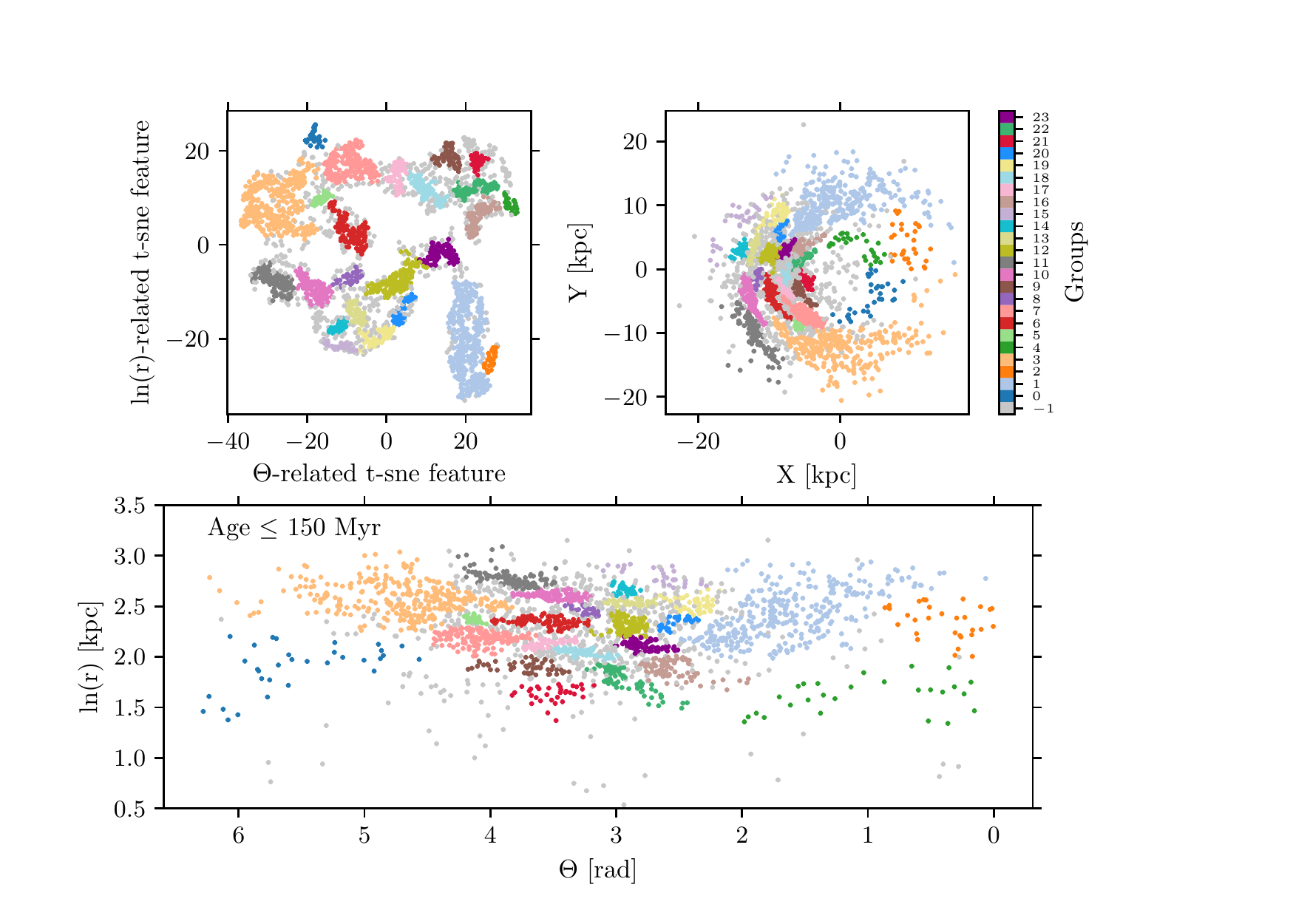}
    \caption{Groups of Cepheids identified by \texttt{HDBSCAN} in the \texttt{t-SNE} space (top left panel). The same groups are presented in the $\theta, ln r$ space (bottom panel), where $\theta$ is the Galactocentric azimuth and ln\,$r$ the logarithm of the Galactocentric radius (corrected from the warp), and in the Galactic plane (top right panel). All the Cepheids shown in this figure are younger than 150\,Myr.}
    \label{fig:single_age_150}
\end{figure*}

 Coming back to real data, the top-left panel in Fig.~\ref{fig:single_age_150} shows how Cepheids younger than 150\,Myr are distributed in the \texttt{t-SNE} space. In this plot, the color-coding indicates groups identified by \texttt{HDBSCAN}, a clustering algorithm using unsupervised learning to identify clusters in a distribution of data points \citep{Campello2015,McInnes2017}. \texttt{HDBSCAN} was run with hyperparameters imposing a minimum of 5 groups, well below the number of clusters actually found, a minimum of 20 members per group in order to avoid spurious detections of tiny groups, and assuming Euclidean distances between individual points in the embedded space.

\par Using the same color-coding, the bottom panel of Fig.~\ref{fig:single_age_150} shows the Cepheids in the ($\theta, ln r$) space. The groups identified by \texttt{t-SNE+HDBSCAN} form narrow, linear sequences in this plane, as is expected under the common assumption that spiral arms follow a logarithmic spiral. The top-right panel of Fig.~\ref{fig:single_age_150}, showing the spatial distribution of the identified groups in the Milky Way plane suggests that each group forms indeed a section of a given spiral arm. A comparison with the spiral structure obtained via other tracers (see Sect.~\ref{sect:model_comp}) confirms that our method allows us to trace the Milky Way spiral arms using Cepheids.

\par As already mentioned, \texttt{t-SNE} does not perform well in the case of sparse data, and the large groups (1,3) gathering distant Cepheids reflect this weakness. They do not trace reliable spatial structures, but emerge only because the search of pulsating variable stars is still largely incomplete (and their classification uncertain) at large distances in the disk, especially toward regions of high extinction like the far side of the disk. Since they do not correspond to real features, these groups will not be discussed further in the paper. Similarly, a few isolated Cepheids in the outer disk are attributed to likely unreliable groups, for instance, to groups 2 or 15.

\par From tests (see Fig.~\ref{fig:age_cuts}) where the sample of Cepheids considered is restricted to those younger than 100, 120, 150, and 250\,Myr, respectively, we draw several conclusions:
\begin{enumerate}[nosep]
    \item Whatever the age cut, the groups identified have the same morphology in the \texttt{t-SNE} space, translating into similar spiral arms in the Galactic plane. Increasing the age limit, hence the number of stars, enables us to split the larger groups into subgroups. 
    \item Increasing the age limit also enables us to better identify spiral features toward the outer disk. This is not a surprise since, for instance, \citet{Skowron2019a} already mentioned that younger Cepheids are observed preferentially in the inner disk. Such a trend is counter-intuitive in the context of an inside-out formation of the disk \citep{Matteucci1989}, it is in fact a combined manifestation of the Milky Way's radial metallicity gradient \citep[e.g.][]{Lemasle2008}, where stellar metallicities decrease from the inner to the outer disk, together with the metallicity-dependence of the Cepheids' instability strip \citep[IS, e.g., ][]{Fiorentino2013,DeSomma2020b}, where the age at which a star of a given mass reaches the IS (or possibly does not even cross it) depends on its metallicity.
    \item If the age limit is set too high, features start to blur again.\\
\end{enumerate}

\par In the rest of the paper, we work with a sample restricted to Cepheids younger than 150\,Myr (figures regarding the samples with different age limits are provided in Appendix \ref{App:spiral_age_cuts}). The arbitrary selection of 150\,Myr as an age limit is a compromise in order to identify a good number of spiral features, bearing in mind the earlier discussion on age in Sect.~\ref{sect:age_discussion}.
\par The groups identified here lead us to characterize spiral arms by segments. In the next subsections, we put such a definition in context, we provide the characteristics of each individual segment, we compare our spiral pattern to some of the most commonly used spiral models, and we investigate the age distribution of Cepheids within individual segments.

\subsection{Defining spiral arms by segments}

\par It is not entirely clear whether the definition of spiral arms by segment is a consequence of the algorithm employed (\texttt{t-SNE} focuses on local similarities in the data), of inhomogeneous completeness of the Cepheid data, or simply a natural outcome of the mechanisms driving the formation of spiral arms in the Milky Way. \citet{Poggio2021} mention that a good fraction of the clumpiness they see in their data (possibly even some larger-scale structures) are caused by foreground extinction. \citet{Zari2021} note, however, that some low-density features are not located in regions of strong extinction and are detected in the spatial distributions of many young tracers, a possible indication that those are not artifacts in the data. It is plausible that young stellar tracers have a clumpy distribution, either because they trace the clumpiness of giant molecular clouds (assuming a stationary spiral pattern), or because star formation is associated with the kinematics and the density distribution across spiral arms (if one considers instead that they are the aftermath of a transient phenomenon).  

\par Studies of external galaxies have also shown that spiral arms are not necessarily homogeneous structures, but can present under-/over-densities, or even be defined by contiguous segments in the most pronounced cases \citep[e.g.,][]{Chernin1999,Kendall2011,Honig2015}. Similarly, \citet{Reid2019} introduced kinks in their logarithmic spiral arms to better fit the spatial distribution of their tracers. Spiral segments are also a natural outcome of theoretical models \citep[e.g.,][]{Grand2012,Donghia2013,Melnik2013} and it has been proposed that they are the response of the stellar disk to the growth of overdensities corotating with the disk \citep[see e.g., ][for a detailed discussion on these topics]{Sellwood2021}. 

\par \citet{Udalski2018} mention that OGLE can detect a P=3d Cepheid at $\sim$20\,kpc from the Sun, even with an extinction reaching A$_{I}\approx$4\,mag, and estimate the completeness of the OGLE survey to be on the order of 90\% for classical Cepheids. It seems then reasonable to discard a significant incompleteness of the Cepheids' catalogs. Another aspect to consider is the number of Cepheids. Their progenitors, late B-type stars, are not extremely numerous given the structure of the IMF, and the brevity of the Cepheid phase, a few tens to a few hundreds megayears (depending on their mass and metallicity) makes them rare objects and, as such, likely not the best tool to discriminate between a multiarm and a flocculent Milky Way.
\par The algorithm developed by \citet{Veselova2020} relies only on the tracers considered to determine the properties of spiral arms/segments, without any assumption on the total number of segments or the membership of a given tracer in a specific segment. Using a sample of nearby Cepheids, they found seven spiral segments using the youngest part of the sample and eight using the oldest Cepheids.

\subsection{Parameters of spiral segments traced by Cepheids}

\par In order to determine the properties of a given individual segment, we fit a linear relation in the ($\theta$, ln\,$r$) space through all group members identified by \texttt{t-SNE+HDBSCAN}:
\begin{equation}
   ln\,r = a \times \theta + b,  
\end{equation}
where $a$ is the slope and $b$ the intercept. From the slope, we derive the pitch angle. The minimal and maximal Galactocentric azimuths covered by a given group are also recorded. The midpoint of these two values is used as the reference angle, and the corresponding radius, calculated using the fitted linear relation, is the reference (logarithm of the) radius. These values, listed in Table~\ref{tab:params_segments_150}, are then used to trace the spiral segments displayed in Fig.~\ref{fig:plot_spiral_segments_150}. As can be seen on the figure or in the tabulated data, several segments are located at a similar reference radius and can be interpreted as different sections of the same spiral arm. 

\begin{table*}
\footnotesize
\centering
\caption{Characteristics of individual segments of spiral arms as identified by \texttt{t-SNE+HDBSCAN} for Cepheids younger than 150\,Myr.}
\label{tab:params_segments_150}
\begin{tabular}{ccccccccccc}
\hline\hline
Group &  Slope & $\sigma_{slope}$ & Intercept & $\sigma_{intercept}$ & Ref. angle & Ref. radius & ln(Ref. radius) & Pitch angle & Min. angle & Max. angle \\
      & (kpc.rad$^{-1}$) & (kpc.rad$^{-1}$) &      (kpc) & (kpc) &    (rad) &      (kpc) &         (kpc) &        (rad) &     (rad) &     (rad) \\
\hline
  {[}4] & -0.090 &       0.050 &     1.705 &       0.060 &     1.068 &      4.997 &         1.609 &        0.090 &     0.154 &     1.982 \\
 {[}21] & -0.082 &       0.087 &     1.922 &       0.304 &     3.502 &      5.129 &         1.635 &        0.082 &     3.177 &     3.827 \\
 {[}22] &  0.575 &       0.042 &     0.087 &       0.124 &     2.835 &      5.568 &         1.717 &       -0.522 &     2.436 &     3.233 \\
 {[}16] &  0.187 &       0.047 &     1.399 &       0.121 &     2.379 &      6.321 &         1.844 &       -0.185 &     1.950 &     2.807 \\
  {[}9] &  0.052 &       0.032 &     1.707 &       0.120 &     3.776 &      6.708 &         1.903 &       -0.052 &     3.375 &     4.177 \\
  {[}0] & -0.255 &       0.073 &     3.304 &       0.407 &     5.424 &      6.827 &         1.921 &        0.250 &     4.566 &     6.282 \\
 {[}18] &  0.163 &       0.019 &     1.512 &       0.061 &     3.253 &      7.708 &         2.042 &       -0.162 &     2.980 &     3.526 \\
 {[}23] &  0.145 &       0.038 &     1.711 &       0.107 &     2.759 &      8.257 &         2.111 &       -0.144 &     2.511 &     3.007 \\
 {[}17] & -0.212 &       0.026 &     2.882 &       0.093 &     3.527 &      8.451 &         2.134 &        0.209 &     3.317 &     3.736 \\
  {[}7] &  0.039 &       0.025 &     2.031 &       0.101 &     4.044 &      8.924 &         2.189 &       -0.039 &     3.639 &     4.450 \\
 {[}12] & -0.023 &       0.062 &     2.381 &       0.181 &     2.945 &     10.107 &         2.313 &        0.023 &     2.691 &     3.198 \\
  {[}6] &  0.048 &       0.020 &     2.176 &       0.072 &     3.602 &     10.474 &         2.349 &       -0.048 &     3.220 &     3.984 \\
  {[}5] &  0.357 &       0.124 &     0.900 &       0.511 &     4.120 &     10.705 &         2.371 &       -0.343 &     4.031 &     4.208 \\
  {[}8] &  0.289 &       0.072 &     1.520 &       0.235 &     3.274 &     11.777 &         2.466 &       -0.281 &     3.141 &     3.406 \\
 {[}13] & -0.006 &       0.029 &     2.554 &       0.082 &     2.889 &     12.637 &         2.537 &        0.006 &     2.684 &     3.094 \\
 {[}10] &  0.041 &       0.020 &     2.460 &       0.069 &     3.516 &     13.520 &         2.604 &       -0.041 &     3.207 &     3.826 \\
 {[}14] &  0.290 &       0.121 &     1.814 &       0.355 &     2.919 &     14.302 &         2.660 &       -0.282 &     2.806 &     3.031 \\
 {[}11] &  0.212 &       0.033 &     1.962 &       0.125 &     3.868 &     16.150 &         2.782 &       -0.209 &     3.480 &     4.256 \\
\hline
\end{tabular}
\tablefoot{The slope and intercept have been computed in the ($\theta, ln r$) space. The reference angle and radius have no physical meaning, they are simply selected as the midpoint of the Galactocentric azimuth range covered by a given group.}
\end{table*}

\subsection{Comparison with spiral arms models}
\label{sect:model_comp}

\par
\begin{figure*}[!ht]
    \centering
    \includegraphics[width=\linewidth]{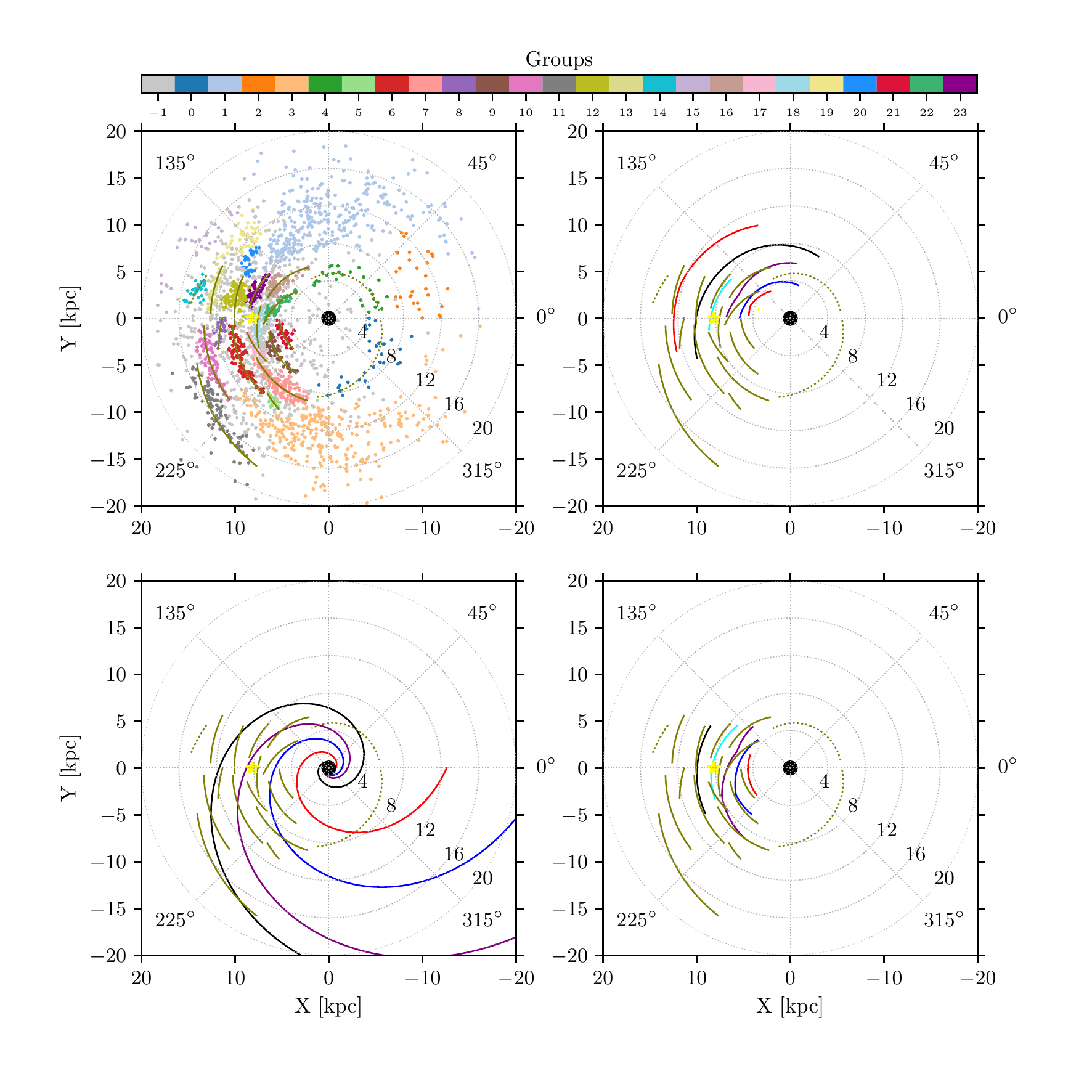}
    \caption{Comparison with previous models. 
    {\it Top left:} Spiral segments (olive) over-plotted on Cepheids identified as members of a group. Groups are color-coded as in Fig.~\ref{fig:single_age_150}. {\it Top right:} Spiral segments and the model of \citet{Reid2019}. The spiral arms delineated by \citet{Reid2019} are over-plotted using the same color-coding as in the original paper. {\it Bottom left:} Spiral segments and the model of \citet{Levine2006b}, with the same color-coding as in \citet{Reid2019}. {\it Bottom right:} Spiral segments and the model of \citet{Hou2021}, with the same color-coding as in \citet{Reid2019}.}
    \label{fig:plot_spiral_segments_150}
\end{figure*}

 Fig.~\ref{fig:plot_spiral_segments_150} displays the spiral segment we derived for Cepheids younger than 150\, Myr over-plotted on our data or on various spiral models, namely those of \citet{Levine2006b}, \citet{Reid2019}, and \citet{Hou2021}. Figures similar to Fig.~\ref{fig:plot_spiral_segments_150} for the other age ranges are provided in appendix \ref{App:spiral_arms}.

\par Several large groups (1, pale blue; 2, orange, 3, pale orange) are not resolved by our algorithm. They are located at large distances from the Sun in the first and in the fourth quadrant. In addition to sparser data in these regions, it is possible that even only slightly larger uncertainties on the distances, and/or a larger number of contaminants, blur the spiral arms signal. Having no physical meaning, these groups are not considered further. 

\par Two other groups (0, dark blue;  4, green) trace long (several kiloparsecs) spiral segments and are defined by a relatively small number of Cepheids. It could be that the stars were connected simply from the lack of further data, but is possible that some of them actually trace real features. They are located in the far side of the (inner) disk. \citet{Minniti2021} have reported that these Cepheids are compatible with a two-arm model (Perseus and Sct-Cen) branching out into four arms for $R_{G}\gtrapprox$~5--6\,kpc.

\par Group 21 (red) seems to prolong the Sct-Cen arm from the model by \citet{Reid2019}. However, it better follows the Norma arm as charted by \citet{Hou2021}. Group 9 (brown) is also an excellent match to the Sct-Cen arm by \citet{Hou2021}, while it would appear to be more of a continuation of the Sgr-Car as defined by \citet{Reid2019}. Group 22 (green) seems to bridge the Sct-Cen and the Sgr-Car spiral arms, according to both models by \citet{Reid2019} and \citet{Hou2021}. It may be that here the algorithm is  unable to separate two closely parallel segments. If so, this group should actually be split in two parts, which would then follow the Sct-Cen and the Sgr-Car spiral arm, respectively. 

\par Group 16 (light brown) is an almost exact match to the Sgr-Car arm from \citet{Reid2019}, and remains in reasonable agreement with the definition of this arm by \citet{Hou2021}. Group 18 (light cyan) constitutes a plausible continuation of the Sgr-Car arm from \citet{Reid2019}, and indeed it partially overlaps this arm in the model by \citet{Hou2021}. Closer to the Sun, this segment, however, approaches the Local arm. 

\par Group 17 (light pink) continues the Local arm as traced by \citet{Reid2019}, but it would reach the Sgr-Car arm if the latter were prolonged with the same pitch angle. And indeed this feature overlaps the Sgr-Car arm in the model by \citet{Hou2021}. Group 23 (violet) overlaps very well with the Local arm when compared to both models. Group 7 (coral pink) is located at the same Galactocentric distance as the Local arm ($\sim$8\,kpc), but it could equally be an extension of the Sgr-Car or of the Local arm, especially if we adopt the location proposed by \citet{Hou2021} for the latter.

\par Both group 12 (khaki) and group 6 (red) are good matches to the Perseus arm as defined by \citet{Reid2019} and especially \citet{Hou2021}. Group 5 (pale green) further extends the Perseus arm toward the fourth quadrant.

\par
\begin{figure}[!ht]
    \centering
    \includegraphics[width=\linewidth]{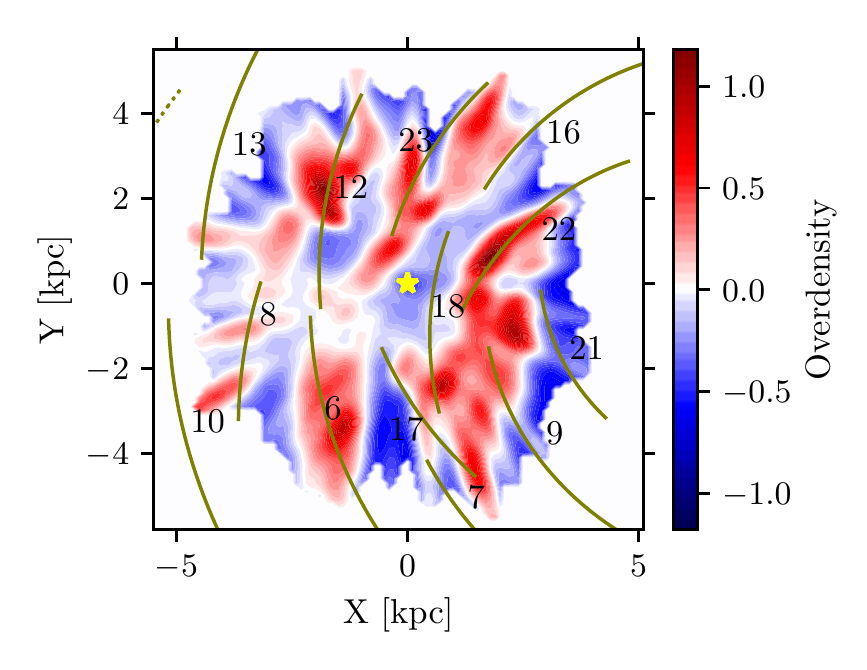}
    \caption{Overdensities in the spatial distribution of upper main sequence stars \citep{Poggio2021}, based on a local density scale length of 0.3\,kpc. The Sun is represented by the yellow star at (0,0). Our spiral segments (olive) are over-plotted.}
    \label{fig:overplot_poggio_150}
\end{figure}

 Similar conclusions can be reached when comparing our segments to the spatial overdensities reported by \citet{Poggio2021} in their sample of upper main sequence stars, thus comparable objects as those used by \citet{Hou2021} to trace spiral arms. The outcome of this exercise is displayed in Fig.~\ref{fig:overplot_poggio_150}. It suggests that groups 18 (light cyan), 17 (light pink) and 7 (coral pink) are all related to the Sgr-Car arm, while group 23 (violet) is the Local arm. It remains unclear whether group 6 (red) also belongs to the Local arm, as suggested by \citet{Poggio2021}, or whether it is a continuation of the Perseus arm, as suggested by Fig.~\ref{fig:plot_spiral_segments_150} from its pitch angle as well as from the good match with the models by \citet{Reid2019} and \citet{Hou2021}. In the latter case, the Local arm might end close to the present location of the Sun. 

\par Our results underline the difficulty to unravel the Milky Way spiral structure in the Solar vicinity\footnote{We note in passing that 16 Cepheids within 1\,kpc from the Sun have been discarded, most of them because they did not fulfill the fractional parallax uncertainty criterion. Including them would increase the amount of nearby Cepheids by 30\%.}. They confirm that the Local arm is not a short segment or a spur emanating from another arm \citep{vandeHulst1954} but an independent, elongated structure \citep{Reid2019,Xu2021}, extending at least from $\approx$135$^{\circ}$ to $\approx$180$^{\circ}$, and possibly approaching the Sgr-Car arm at Galactocentric azimuths $\approx$200$^{\circ}$. 

\par Group 13 (pale khaki) is an excellent match to the Outer arm as sketched by \citet{Reid2019}, and group 10 (pink) seems to constitute its natural continuation, although it is located at a somewhat larger Galactocentric distance than modeled by \citet{Reid2019}. Group 14 (cyan) is identified as a potential spur extending out of the Outer arm in the anticenter direction. {Group 8 appears as an isolated segment in between the Perseus and the Outer arms.}

\par Groups 19 (pale yellow) and 20 (blue) are not a reliable feature. In the range 135$^\circ$--150$^\circ$, at least a fraction of the Cepheids attributed to these groups should arguably belong to group 12 (khaki) or group 13 (pale khaki). In this region, the algorithm is strongly affected by a shadow cone\footnote{where the detection of targets is hampered beyond nearby regions with strong extinction} already reported by \citet[][their Fig.~1]{Poggio2021}, which is also quite obvious in our data (another shadow cone splits group 1 and probably prevents the algorithm to identify groups in these distant regions of the first quadrant). The reality of group 15 (pale violet) cannot be assessed due to the paucity of Cepheids in the far outer disk, Group 11 (gray) can be attributed to the Outer arm only if we accept a drastic change in its pitch angle. This feature agrees quite well with the model by \citet{Levine2006b}, where it represents, however, the Perseus arm, as already noted by \citet{Poggio2021}. The data are too sparse to draw firm conclusions, and especially to trust that this feature indeed extends up to $\sim$250$^\circ$.

\par We also compared our findings to the study by \citet{Veselova2020}. It relies on a much smaller sample of Cepheids \citep[636, from][]{Melnik2015} with overall less accurate distances (but well-constrained radial velocities). They consider segments as a section of a logarithmic spiral. The membership of Cepheids in segments and the properties of each segment are determined simultaneously. Matching spiral segments in \citet{Veselova2020} to our groups was carried out via a visual inspection of Fig.~\ref{fig:plot_spiral_segments_150} in this study and Fig.~5 in theirs. When possible, we included in the comparison some of our segments lying outside the (small) range of Galactocentric azimuths encompassed by the study of \citet{Veselova2020}. We note that group 19 (red), which we associate with their Sagittarius-2 segment, only seems to have a limited length in our study (but one could argue that the small spur visible at the extremity of group 4 (green) might be a continuation of this feature). 
\par In Table~\ref{tab:comp_Vese}, we compare the Galactocentric distance of the spiral arms in their study and in ours. The results are in very good agreement until the (first) Outer arm. We note that the two outermost features are only defined by a select number of stars in \citet{Veselova2020}, while our comparison groups contain much more stars, from which, however, only a few overlap with the spatial extent of the sample by \citet{Veselova2020}. The consistency between the two studies suggests that our detection of spiral arms is robust, even outside the comparison region. We also find a fairly good agreement with the groups of Cepheids identified by \citet{Genovali2014}, without considerations on age. The identification was based on a clustering algorithm \citep[Path Linkage Criterion,][]{Battinelli1991} already applied to Galactic Cepheids by \citet{Ivanov2008}, and on a stellar density threshold between the candidate groups and their immediate neighborhood. 

\begin{table*}[!ht]
\centering
\small
\caption{Comparison of the Galactocentric distances of spiral arms in this study with the Galactocentric distances of spiral arms in \citet{Veselova2020} (V20) and with the Galactocentric distances of Cepheids groups identified by \citet{Genovali2014} (G14).}
\label{tab:comp_Vese}
\begin{tabular}{rcccrcr}
\hline\hline
          Arm &  G14  & Group(s) & V20 "Young" & V20 "Old" & This study & Group(s)  \\
              & (kpc) &          &   (kpc)     &   (kpc)   &    (kpc)   &        \\
\hline
Scutum        &       &          &        5.94 &      6.07 &       5.35 & 21,22  \\
Sagittarius-1 &  6.41 &     I,II &        6.74 &      6.79 &       6.62 & 0,9,16 \\
Sagittarius-2 &  7.45 &  III, IV &        7.41 &      7.52 &       7.71 & 18     \\
Local         &  8.10 &        V &        8.31 &      8.19 &       8.35 & 17,23  \\
              &  8.99 &       VI &             &           &       8.92 & 7      \\
Perseus       &  9.54 & VII, VIII&       10.13 &      9.91 &      10.11 & 12     \\
              & 10.13 &    IX, X &             &           &      10.59 & 5,6    \\   
Outer-1       &       &          &       12.59 &     12.49 &      13.08 & 10,13     \\
Outer-1a      &       &          &             &     14.68 &      14.30 & 14     \\
Outer-2       &       &          &       16.61 &     16.79 &      16.15 & 11     \\
\hline
\end{tabular}
\tablefoot{In \citet{Veselova2020}, the Cepheids are split into an old (P$\leq$9d) and a young (P$\geq$9d) sample. Heliocentric distances of the spiral arms of \citet{Veselova2020} are taken from their Table~1, and converted into Galactocentric distances by adding 7.33\,kpc and multiplying the outcome by a correction factor of 1.117, as indicated by those authors. The Galactocentric distances of the spiral arms in this study are the average of the reference radius of individual groups (see Table~\ref{tab:params_segments_150}). Groups representative of a given spiral arm have been selected by comparing Fig.~\ref{fig:plot_spiral_segments_150} in this study and Fig.~5 in \citet{Veselova2020}. For the Cepheids groups identified by \citet{Genovali2014}, we use directly the tabulated Galactocentric distances or their average value if we combine two of their groups. We note that \citet{Genovali2014} used 7.94\,kpc \citep{Groenewegen2008b} as the Galactocentric distance of the Sun.}
\end{table*}

\subsection{Age distribution of Cepheids across spiral segments}

\par To investigate the age distribution of Cepheids in our individual spiral segments, we use the data tabulated in Table~\ref{tab:params_segments_150} to rotate each data point by the pitch angle value around the reference point in the ($\theta$, ln\,$r$) space. It then becomes easy to derive the (logarithmic) distance of a given Cepheid from the (logarithmic) reference radius, and to convert it into real spatial distances. We then plot the ages of the Cepheids attributed to a given group against their distance to the reference radius. Age gradients across individual segments are shown in Fig.~\ref{fig:age_gradient_a}, together with linear fits to the data.

\par The period-age relations by \citet{Bono2005} and \citet{Anderson2014b} do not allow us to derive uncertainties on the age of individual Cepheids. The standard deviation of the period-age relation by \citet{Bono2005}, which mainly reflects the finite width of the instability strip, can be used as a (loose) proxy for the 1-$\sigma$ uncertainty on age. We find that it can reach 15\,Myr for nonrotating Cepheids. Including rotation in models increases the ages of Cepheids by 50 to
100\%, depending on the amount of rotation and the period of the Cepheid \citep{Anderson2014b}. Finally, the helium and metal contents, a possible core convective overshooting during the core H-burning stage, and the mass loss efficiency (mostly) during the red giant branch phase, all affect the model predictions of the Cepheids' individual ages \citep[see e.g.,][]{DeSomma2020b}. The theoretical period-age relations are provided only for selected values of these parameters. The true values of these physical quantities likely vary from star to star, and they are anyway not available to us at this time, a fortiori for large samples. It is important for the current analysis to mention these caveats, but they should not hide the fact that Cepheids are among the stars with the best age determinations.

\par Beyond the uncertainties on the Cepheids' ages already mentioned above, and the underlying assumption that our spiral segments can be defined by sections of a logarithmic spiral arm, we note that the birth location of the Cepheids is still unknown. \citet{Medina2021} report only a relatively small number of Cepheids confidently associated with open clusters, which may indicate that Cepheids are born elsewhere, or alternatively that their birth cluster or association dissolved quickly. The dynamical evolution of Cepheids in star clusters has been investigated theoretically by \citet{Dinnbier2022}. Moreover, the recent discovery of numerous spurs and feathers \citep[e.g.,][and references therein]{Kuhn2021,Veena2021}, sometimes extending quite far away from the estimated locus of the corresponding spiral arm, hints at a more complicated picture.

\par Still, our analysis already provides a gross estimate of the age gradient across spiral arms. Numerous studies \citep[e.g.,][to quote only a few recent ones]{Shabani2018,Castro-Ginard2021} have searched for age gradients in our or in external galaxies in order to test the spiral density wave theory \citep[see e.g.,][]{Dobbs2010}, but the majority of them simply report their absence or their detection \citep[but see][]{Vallee2020}. The values we obtain range from 0 to $\approx$15\,Myr\,kpc$^{-1}$. They agree quite well with the age gradients of 12$\pm$2\,Myr\,kpc$^{-1}$ reported by \citet{Vallee2021}. We note in passing that given the age limit we set at 150\,Myr, a few age distributions for individual segments are potentially truncated at higher ages and may provide unreliable values.

\section{Spiral arms and abundance gradients}
\label{sect:gradients}

\par In this section, we investigate how the spiral arms may impact radial and azimuthal\footnote{By azimuthal metallicity gradient, we mean the variation of the metallicity with the Galactocentric azimuth, within a given radial annulus at fixed Galactocentric distance.} metallicity gradients, using literature values from \citet{Genovali2014} and references therein. We emphasize that the conclusions drawn in this section are only tentative since we list below a number of important caveats that should be kept in mind.

\subsection{Caveats}
\label{sect:Caveats}

\par The first caveat we want to mention is that the spectroscopic analysis of Cepheids is not immune from (phase-dependent) NLTE effects \citep[see the series of papers by][for instance]{Vasilyev2017,Vasilyev2018,Vasilyev2019}. NLTE effects are more important for more massive Cepheids, which are also the longer period, younger ones that are presumed to be the best tracers of spiral arms.\newline
\indent The catalog of Cepheid metallicities by \citet{Genovali2014} contains several tens of Cepheids analyzed in their paper, using the same method as for the stars in \citet{Lemasle2007,Lemasle2008,Romaniello2008,Pedicelli2010,Genovali2013} that they also included in their catalog. They complemented those studies with literature data from other groups, namely \citet{Sziladi2007,Yong2006,Luck2011a,Luck2011b}, which were rescaled to take systematic differences between these studies into account. Although these differences are $\lesssim$0.1\,dex (with the only exception of \citet{Yong2006} who report notably lower metallicities for the stars in common), we are still dealing with inhomogeneous data.\newline
\indent The strongest caveat is related to line depth ratios \citep[LDR, see][]{Kovtyukh2000,Kovtyukh2007,Proxauf2018}, the traditional method to determine the effective temperature of Cepheids, prior to a canonical spectroscopic analysis. Even though the temperature scale derived from line depth ratios is quite precise (better than 100\,K), several concerns have arisen regarding the accuracy of the scale, and possible systematic errors depending on the metallicity of the star and the phase of the observation have been suggested \citep[e.g.,][]{Mancino2021}. Such potential systematics have motivated the series of papers initiated in \citet{Lemasle2020}, where we aim at developing an unbiased method for deriving the chemical composition of Cepheids.\newline  
\indent Finally, we note that the spectroscopic data available in the literature contain mostly stars within a few kiloparsecs from the Sun, and, therefore, cover only a fraction of the spiral segments identified in this study, some of them very partially (in terms of their longitudinal extension).

\subsection{The radial gradient of iron}

\begin{figure*}[!ht]
    \centering
    \includegraphics[width=\linewidth]{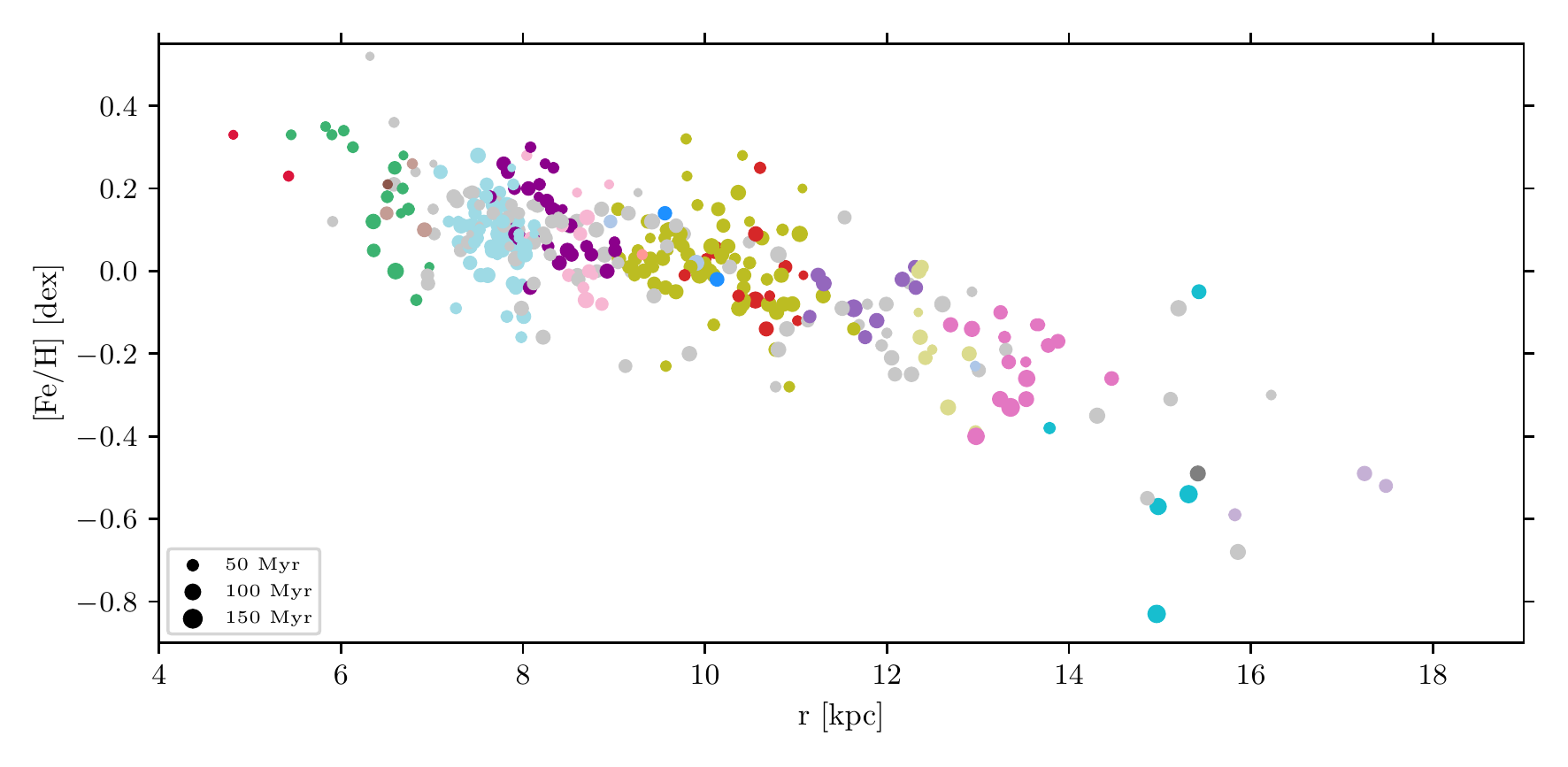}
    \caption{Radial metallicity gradient, where [Fe/H] derived from high-resolution spectroscopic observations \citep{Genovali2014} is plotted against Galactocentric distances $r$ determined using a period-luminosity relation (this study) in the mid-infrared. The different spiral segments are color-coded as in Fig.~\ref{fig:single_age_150}, and the size of the data-points correlates with the age of the Cepheids (larger points for higher ages). As in previous figures, only Cepheids with an age$\leq$150\,Myr are considered.}
    \label{fig:radial_gradient}
\end{figure*}

\par Assuming a logarithmic spiral structure, one could expect that stars located in different spiral arms may overlap in a 2D representation of the radial metallicity gradient (where the Galactocentric azimuth is not considered), as shown in Fig.~\ref{fig:radial_gradient}, For the data we have at hand, this is not the case, but this most likely simply reflects the limited longitudinal extension of the spectroscopic data. A gap at r$\approx$11.5\,kpc indicates a lack of young Cepheids at this radius, but it is already vanishing when considering slightly older Cepheids (see Fig.~\ref{fig:age_cuts}).

\par An interesting feature of Fig.~\ref{fig:radial_gradient} concerns the observed scatter in [Fe/H] at a given radius. It is not clear yet whether this scatter is real or only a consequence of uncertainties in the chemical analysis, but the [Fe/H] scatter for Cepheids attributed to a given segment or for unclassified (potentially inter-arms') Cepheids is on the same order of magnitude. In a companion paper from our group \citep{daSilva2022}, we show that the dispersion around the mean gradient can be reduced by tackling some of the systematics (line list, atomic parameters) affecting abundance estimates.  

\par Given the caveats exposed above, we did not try to investigate whether a simple linear gradient or more complicated features, including, for instance, breaks in the slope, would best represent the data. Such an investigation is possible using modern regression methods, but it would require robust estimates of the accuracy and precision of the abundance determination, which we have set as a goal for the series mentioned earlier, and a larger sample of Cepheids, which will soon become available from WEAVE \citep{Dalton2016} and 4MOST \citep{Chiappini2019,Bensby2019}. For the same reasons, we did not yet consider other elements except iron.

\subsection{The azimuthal gradient of iron}

\par
\begin{figure}[!ht]
    \centering
    \includegraphics[width=\linewidth]{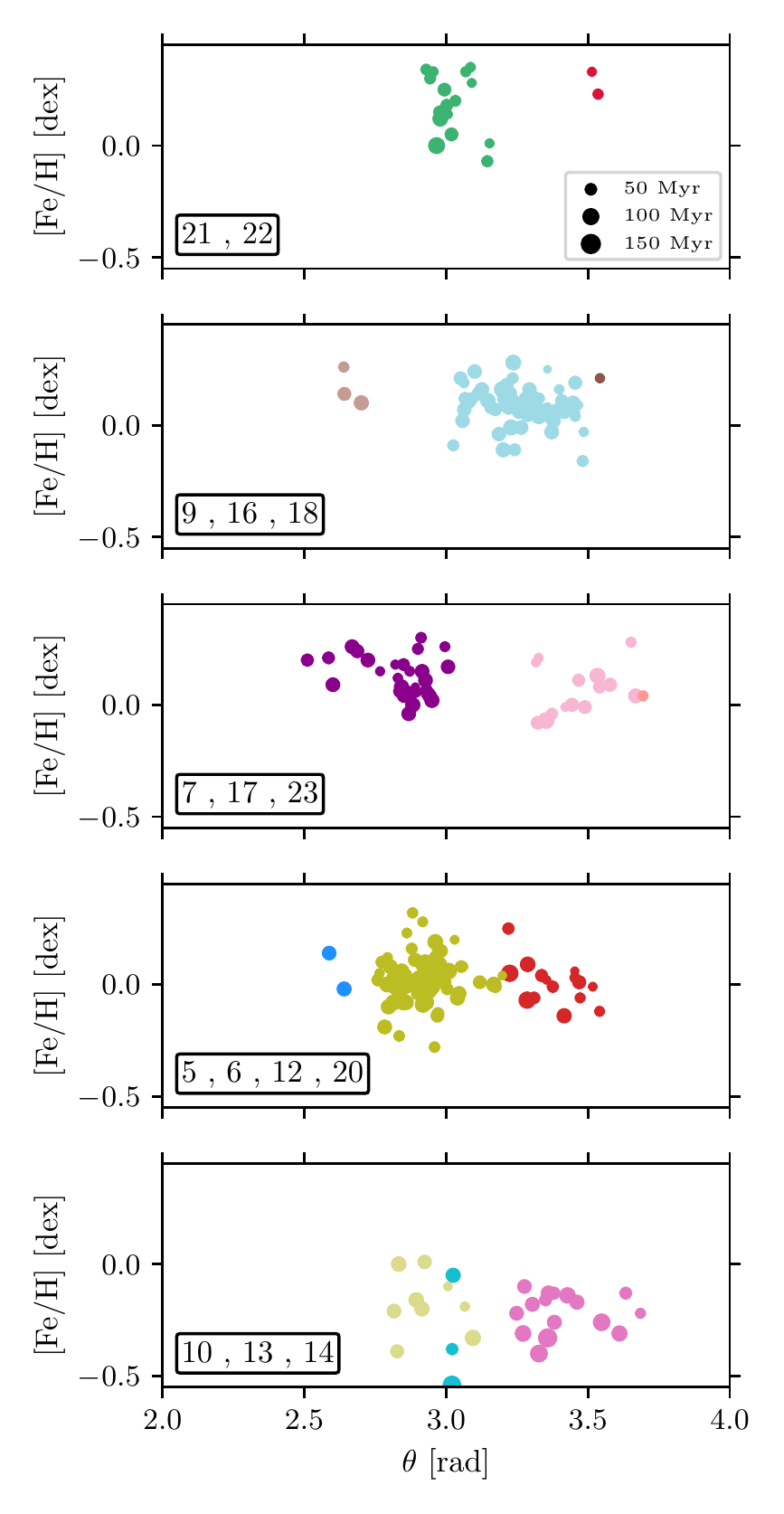}
    \caption{Longitudinal metallicity gradient, where [Fe/H] derived from high-resolution spectroscopic observations \citep{Genovali2014} is plotted against the Galactocentric azimuth $\theta$. The different spiral segments are color-coded as in Fig.~\ref{fig:single_age_150}, and the size of the data-points correlates with the age of the Cepheids (larger points for larger ages). As in previous figures, only Cepheids with an age$\leq$150\,Myr are considered.}
    \label{fig:azimuthal_gradient}
\end{figure}

 Fig.~\ref{fig:azimuthal_gradient} shows the metallicity distribution within spiral segments corresponding to several spiral arms. Keeping the caveats mentioned in Sect.~\ref{sect:Caveats} in mind, the metallicity excursion barely exceeds 0.4\,dex within a given spiral arm, and there are some hints of a metallicity trend with the Galactocentric azimuth (but the range of Galactocentric azimuths covered is relatively small), possibly increasing toward the outer disk. Such results are in line with the conclusions drawn by \citet{Kovtyukh2022}. Excellent precision and accuracy are required since model predictions indicate that the expected effects are modest. For instance, \citet{Spitoni2019} indicate that azimuthal variations in the oxygen abundance gradient are on the order of 0.1\,dex. \citet{Molla2019} reach similar conclusions. Within a given segment or spiral arm, we also see no metallicity trend with age.

%----------  CONCLUSION  ----------

\section{Conclusions}
\label{sect:conc}

\par In this paper we took advantage of updated mid-infrared photometry and of the most complete (to date) catalog of Galactic Cepheids to determine the shape of the Milky Way warp using a Bayesian robust regression method. We have derived the Galactocentric distances of individual Cepheids in a nonwarped Milky Way and we concluded that the warp cannot be responsible for the increased dispersion of abundance gradients in the outer disk.
\par Thanks to a clustering algorithm, we have identified groups of Cepheids in the ($\theta$, ln\,$r$) space, where $\theta$ and $r$ are the Galactocentric azimuth and distance (corrected from the effects of the warp) of individual Cepheids. Assuming different values for the oldest Cepheid considered, we have fit these groups with segments in the ($\theta$, ln\,$r$) space, which translate into portions of spiral arms in the ($\theta$, $r$) space. These groups are consistent with previous studies mapping the density of young tracers in the Solar neighborhood \citep[e.g.,][]{Poggio2021,Zari2021}, or using them to derive explicitly the location of the spiral arms \citep[e.g.,][]{Reid2019,Hou2021}. 

\begin{acknowledgements}
The authors thank the anonymous referee for her/his very useful comments and suggestions. E. K. G., A. K., V. K., H. L., B. L., Z. P. acknowledge the Deutsche Forschungsgemeinschaft (DFG, German Research Foundation) -- Project-ID 138713538 -- SFB 881 ("The Milky Way System", subprojects A05, A08, A11). V. B., R. D. S., M. F. acknowledge the financial support of INAF (Istituto Nazionale di Astrofisica), Osservatorio Astronomico di Roma, ASI (Agenzia Spaziale Italiana) under contract to INAF: ASI 2014-049-R.0 dedicated to SSDC. A. K. acknowledges support from the National Research Foundation (NRF) of South Africa. V. K. is grateful to the Vector-Stiftung at Stuttgart, Germany, for support within the program "2022--Immediate help for Ukrainian refugee scientists" under grant P2022-0064.
\end{acknowledgements}

\bibliographystyle{aa} % style aa.bst
\bibliography{lemasle.bib}

\begin{appendix} %First appendix

\section{The Milky Way warp} 

\subsection{Length of a bow following the warp.}
\label{App:unwarp}

\begin{figure}[!ht]
    \centering
    \includegraphics[width=1.0\linewidth]{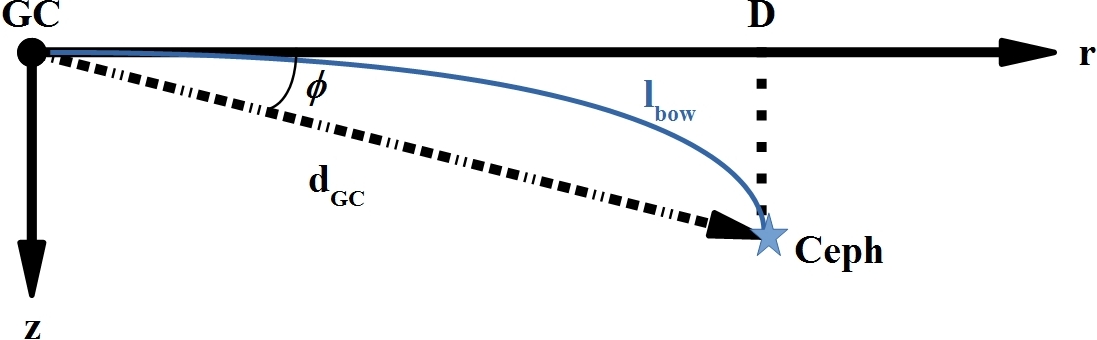}
    \caption{Slice of the (warped) Galactic disk in the Galactocentric azimuth $\theta$ containing a Cepheid, where d$_{\mathrm{GC}}$ is the Galactocentric distance of the Cepheid, D its projection on the Galactic plane, and $\phi$ its Galactocentric latitude.}
    \label{fig:warp_schematic}
\end{figure}

\par We adopt the definition of the warp proposed by \citet{Skowron2019b}, where the Milky Way is warped at radii exceeding a given radius $r_{0}$. At such distances, the warp is given by the equation:
\begin{equation*}
    z(r,\Theta) = z_{0}+(r-r_{0})^{2}\times[z_{1}\,sin(\Theta-\Theta_{1})+z_{2}\,sin(2\,(\Theta-\Theta_{2}))]
\end{equation*}
where $z$ is the vertical distance from the Galactic plane, $r$ is the distance from the Galactic center, and $\Theta$ is the Galactocentric azimuth. The constants $r_0$, $z_0$, $z_{1}$, $z_{2}$, $\Theta_{1}$, $\Theta_{2}$ are radial, vertical, and angular parameters describing the surface.

\par With such a definition, the warp has a parabolic shape, which varies with the Galactocentric azimuth $\Theta$. Indeed, if we set the position of a given Cepheid as $(r_{cep},\Theta_{cep},z_{cep})$ in Galactocentric coordinates, the equation of the warp is of the form:
\begin{equation}
    z=z_0+C\times(r-r_0)^2
\end{equation}
where \begin{equation*}
C=z_{1}\,sin(\Theta_{cep}-\Theta_{1})+z_{2}\,sin(2\,(\Theta_{cep}-\Theta_{2}))      
\end{equation*}
in the vertical plane containing the Galactic center and the Cepheid at the Galactocentric azimuth $\Theta_{cep}$.

\par The length of the bow between the Galactic center and a given Cepheid located at the distance d$_{\mathrm{GC}}$ from the Galactic center is given by the integral equation:
\begin{equation}
    l_{bow}=\int_{0}^{d_{\mathrm{GC}}\cos(\phi)}\sqrt{1+\left(\frac{dz}{dr}\right)^{2}}\,\mathrm{d}r
\end{equation}
where $z=f(r)$ is the function describing the warp and $D=d_{\mathrm{GC}}\cos\phi$ is the Galactocentric distance of the Cepheid projected on the Galactic plane.\medskip

\par Given the definition of the warp given by \citet{Skowron2019b}, we have to integrate:
\begin{equation}
    l_{bow}=\int_{0}^{D}\sqrt{4C^2\left(r-r_0\right)^2+1}\,\mathrm{d}r
\end{equation}
with $D=d_{\mathrm{GC}}\cos\phi$.
After a series of substitutions of variables, it is possible to analytically determine the value of $l_{bow}$. In this work we derived it numerically using \textsc{scipy}.

\newpage
\onecolumn

\subsection{The warp model}
\label{App:warp_model}

\begin{table*}[ht]
\centering
\footnotesize
\caption{Covariance matrix for the Bayesian robust regression of the warp model.}
\label{tab:cov_matrix_warp}
\begin{tabular}{lrrrrrrrr}
\hline\hline
       & \multicolumn{1}{c}{$r_0$} & \multicolumn{1}{c}{$z_0$} & \multicolumn{1}{c}{$z_{1}$} & \multicolumn{1}{c}{$z_{2}$} & \multicolumn{1}{c}{$\Theta_{1}$} & \multicolumn{1}{c}{$\Theta_{2}$} &        \multicolumn{1}{c}{$\sigma$} & \multicolumn{1}{c}{$\nu$} \\
\hline
 $r_0$    &  0.0983427   & -0.000584572 &  0.000173638 &  4.72231e-07 &  0.00244786  & -0.00333678  & -7.20902e-05 & -0.00166704  \\
 $z_0$    & -0.000584572 &  2.41349e-05 & -8.84631e-07 &  1.83132e-07 & -8.14438e-05 &  3.50388e-05 &  1.55684e-06 &  5.25924e-05 \\
 $z_{1}$    &  0.000173638 & -8.84631e-07 &  4.20396e-07 &  3.95521e-08 &  7.20249e-06 &  1.64115e-05 & -1.53575e-07 & -5.55467e-06 \\
 $z_{2}$    &  4.72231e-07 &  1.83132e-07 &  3.95521e-08 &  7.70687e-08 & -4.86087e-06 &  8.99776e-08 &  3.91017e-08 &  1.83853e-06 \\
 $\Theta_{1}$  &  0.00244786  & -8.14438e-05 &  7.20249e-06 & -4.86087e-06 &  0.00112674  &  0.00154508  & -7.66386e-06 & -0.000389352 \\
 $\Theta_{2}$  & -0.00333678  &  3.50388e-05 &  1.64115e-05 &  8.99776e-08 &  0.00154508  &  0.00955416  & -9.70985e-07 & -0.000574705 \\
 $\sigma$ & -7.20902e-05 &  1.55684e-06 & -1.53575e-07 &  3.91017e-08 & -7.66386e-06 & -9.70985e-07 &  1.09651e-05 &  0.000380751 \\
 $\nu$    & -0.00166704  &  5.25924e-05 & -5.55467e-06 &  1.83853e-06 & -0.000389352 & -0.000574705 &  0.000380751 &  0.0283286   \\
\hline
\end{tabular}
\tablefoot{$r_0$, $z_0$, $z_{1}$, $z_{2}$, $\Theta_{1}$, $\Theta_{2}$ are the structural parameters of the warp model, $\sigma$ its standard deviation and $\nu$ its normality parameter.}
\end{table*}

\begin{figure*}[!ht]
  \includegraphics[width=\linewidth]{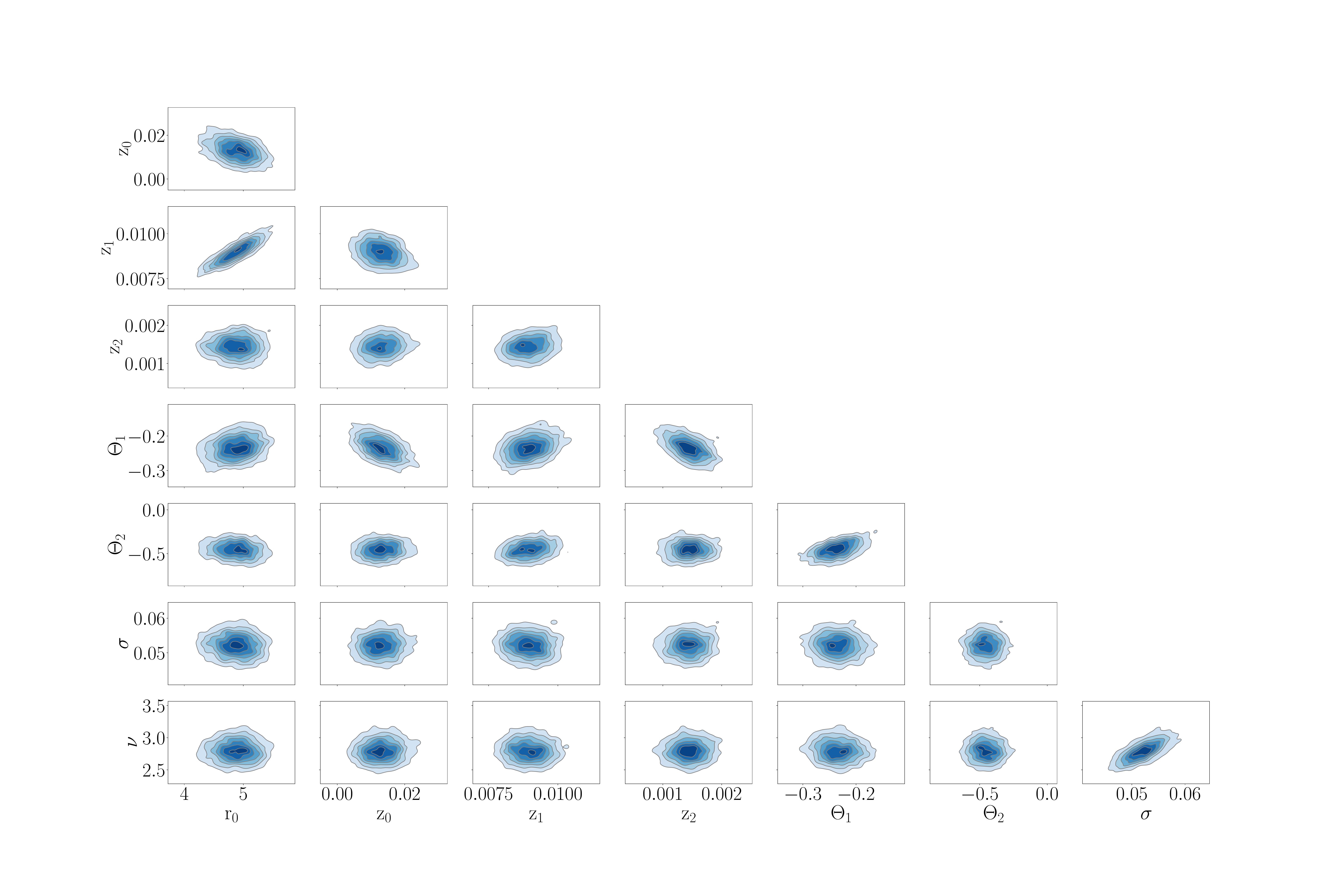}
  \caption{Pairwise correlations of the posterior distributions of the structural parameters $r_0$, $z_0$, $z_{1}$, $z_{2}$, $\Theta_{1}$, $\Theta_{2}$ of the warp model, together with the standard deviation $\sigma$ and the normality parameter $\nu$ of the model.}
  \label{fig:corner_plot_warp}
\end{figure*}  

\clearpage
\begin{figure*}[!ht]
  \includegraphics[width=\linewidth]{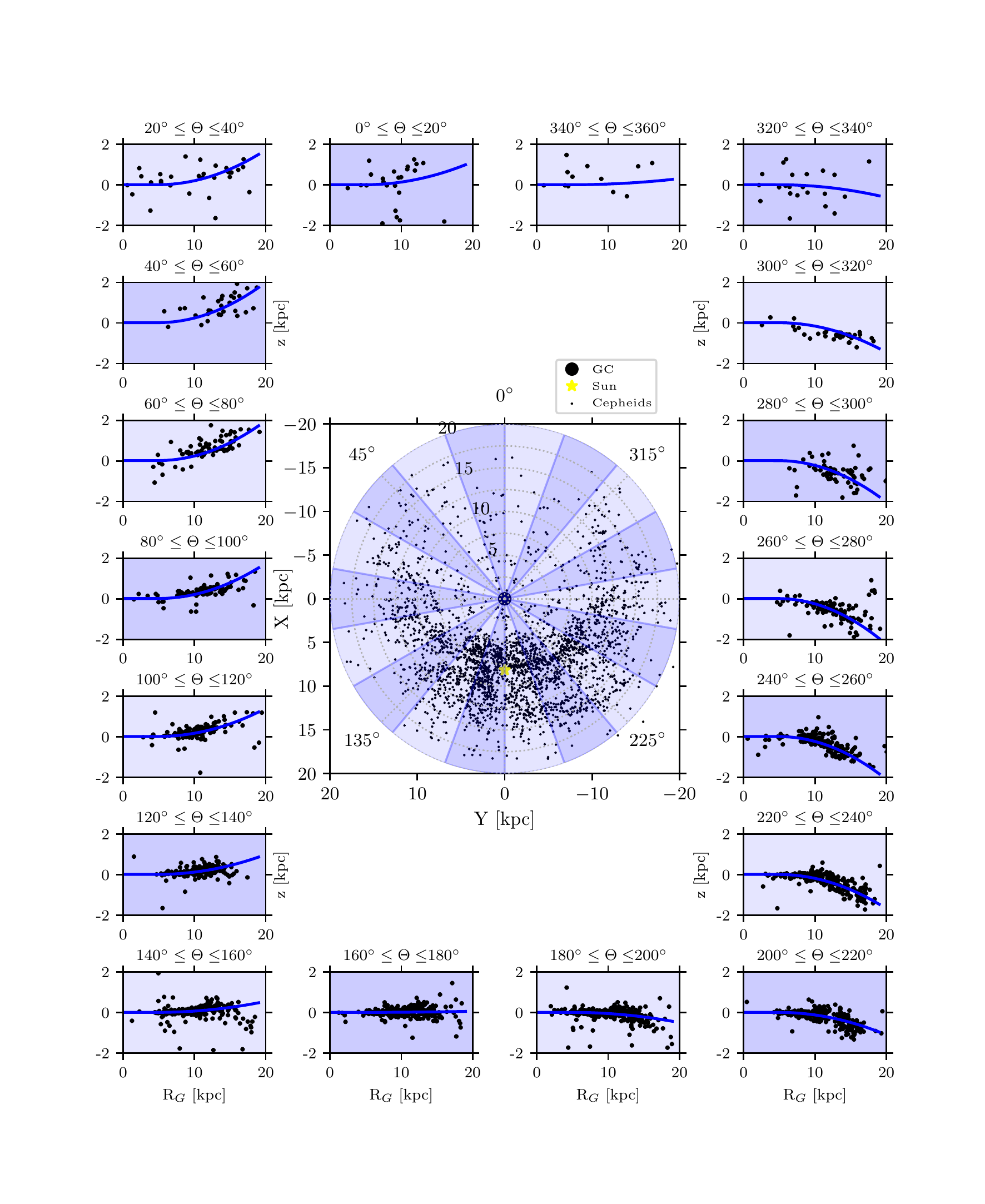}
  \caption{Galactic warp traced by classical Cepheids. The Milky Way disk is divided into 18 slices. The vertical
distribution of Cepheids within each slice is displayed in adjacent panels, showing the distance from the Galactic plane $z$ as a function of the Galactocentric distance $r$. The black dots represent individual Cepheids while the blue line is the warp model, computed with the parameters listed in Table~\ref{tab:params_bayesian} and for the median Galactocentric azimuth of a given slice.}
  \label{fig:warp_sectors}
\end{figure*}  

\newpage
\FloatBarrier
\twocolumn

\section{Testing the algorithm} 
\label{App:test_algo}

\subsection{Creation of mock spiral arms}

\par To create mock spiral arms, we rely on the \citet{Reid2019} model. We selected it over other possible choices \citep[e.g.,][]{Hou2021} because the immediate proximity of the Norma and Sct-Cen arms, especially in the vicinity of the kink in the Norma arm, is very challenging for our algorithm. We want to stress here that only a few Cepheids are currently known at the expected locations of the 3-kpc and Norma arms (see Fig.~\ref{fig:spiral_Reid19}). 
\par We then select N equidistant points along the spiral arms traced by \citet{Reid2019} to populate the spiral arms with mock Cepheids by using N bivariate normal distributions. Such a distribution is defined by its mean and covariance matrix, which are analogous to the mean and variance of a 1D normal distribution. In our case, the means are the (x,y) positions of each of the N reference points and the covariance matrix is $ \left( \begin{array}{cc}
\sigma^{2} & 0 \\
0 & \sigma^{2} \\
\end{array}  \right) $.

\par The values for the parameters of the bivariate distribution are tabulated in Table~\ref{tab:2D-MVN_params}. The values of $\sigma$ are proportional to the widths of the individual spiral arms proposed by \citet{Reid2019}. With these parameters, several ten thousands mock Cepheids are created. For each individual arm, a very small number of them lies outside the angular domain encompassed by the \citet{Reid2019} model, they are discarded for comparison purposes (see Sect.~\ref{sect:Hotelling}). This sharpens the extremities of the mock spiral arms, which would otherwise appear rounded because of stars generated by the very first and very last bivariate distributions. Finally, N$_{spiral}$ mock Cepheids are randomly selected as our spiral arms Cepheids sample.

\begin{table}[ht]
\centering
\caption{Parameters for the creation of mock spiral arms based on the \citet{Reid2019} model.}
\footnotesize
\label{tab:2D-MVN_params}
\begin{tabular}{r|cc|ccc}
\hline\hline
{} & \multirow{2}{*}{$\sigma^{2}$} & \multirow{2}{*}{N} & \multicolumn{3}{c}{Noise parameter}   \\
{} &    {}    & {}  & Ideal & Noisy & Inter arms \\
\hline
3 kpc   &    0.014 &  20 &  0.0 &  0.03 &  0.01 \\
Norma   &    0.011 &  20 &  0.0 &  0.03 &  0.01 \\
Sct-Cen &    0.018 &  20 &  0.0 &  0.03 &  0.01 \\
Sgr-Car &    0.021 &  20 &  0.0 &  0.03 &  0.01 \\
Local   &    0.024 &  50 &  0.0 &  0.03 &  0.01 \\
Perseus &    0.027 &  50 &  0.0 &  0.03 &  0.01 \\
Outer   &    0.050 &  50 &  0.0 &  0.03 &  0.01 \\
\hline
\end{tabular}
\tablefoot{N is the number of centers for bivariate normal distributions with variance $\sigma^{2}$ in their covariance matrix. Noise is the scaling factor for random noise added to the coordinates of the mock stars. Random noise is drawn from an univariate normal distribution of mean 0 and variance 1.}
\end{table}

\subsection{Comparison between the mock and the retrieved sample: An ideal test case}
\label{sect:Hotelling}

\par To quantify the comparison between the original mock data and the retrieved sample, we use Hotelling’s t-Squared \citep[t$^{2}$,][]{Hotelling1931} statistics. It is the multivariate generalization of the Student's t-distribution, and it is specifically useful when comparing two distributions of unequal sizes, mean values and variances. Our null hypothesis is that the original and retrieved samples for a given spiral arm are drawn from the same parent distribution.\\
 
\par
\begin{figure*}[hb]
  \includegraphics[width=\linewidth]{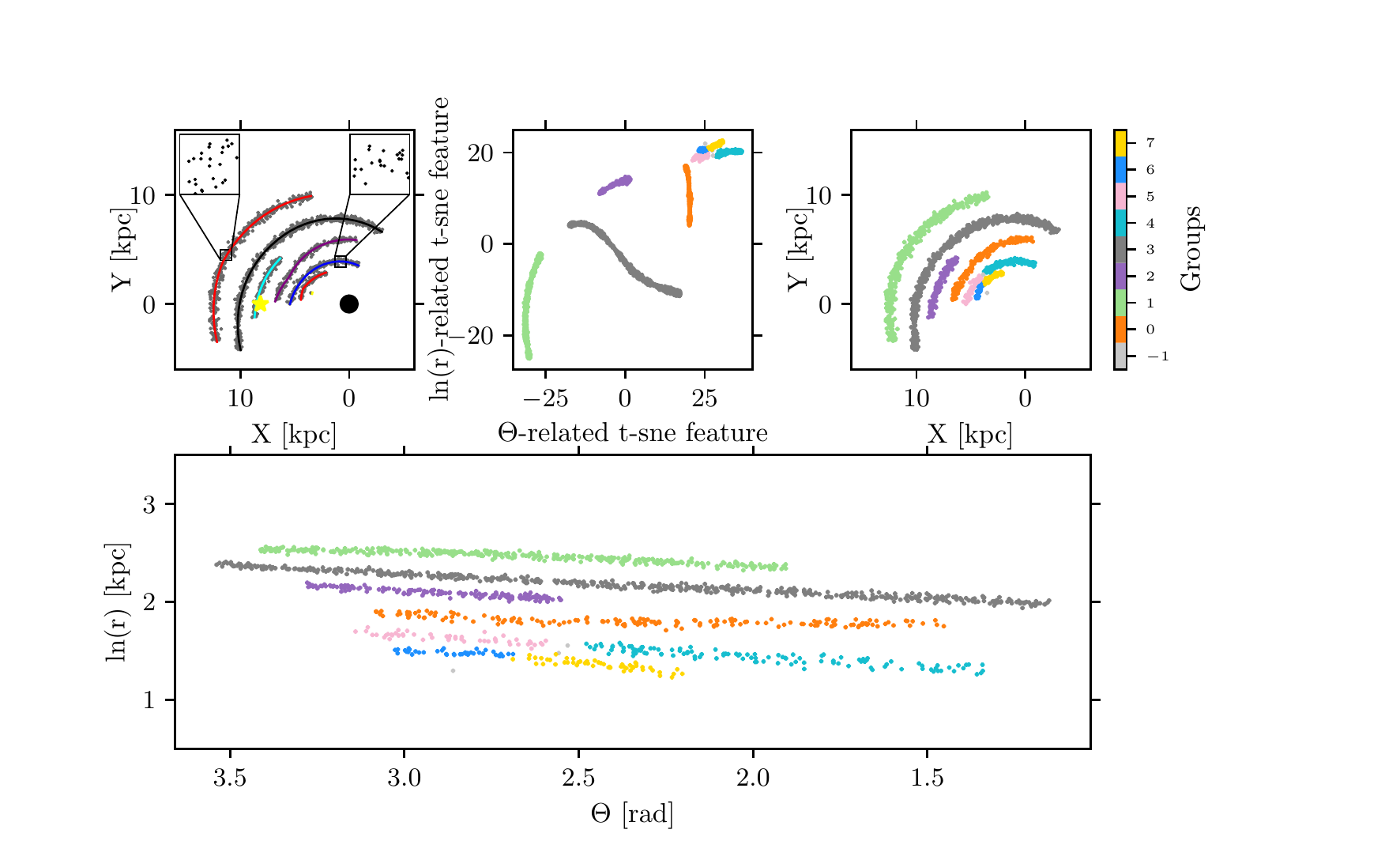}
  \caption{Mock sample of N$_{spiral}$=1500 Cepheids in spiral arms following the model by \citet{Reid2019} (top left panel). Groups of Cepheids identified by HDBSCAN in the t-SNE space (top middle panel). The same groups are presented in the ($\Theta$, ln\,$r$) space (bottom
panel), where $\Theta$ is the Galactocentric azimuth and ln\,$r$ the logarithm of the Galactocentric radius (corrected from the warp), and in the Galactic
plane (top right panel).}
  \label{fig:App1a}
\end{figure*}

\begin{figure*}[hb]
  \sidecaption
  \includegraphics[width=0.5\linewidth]{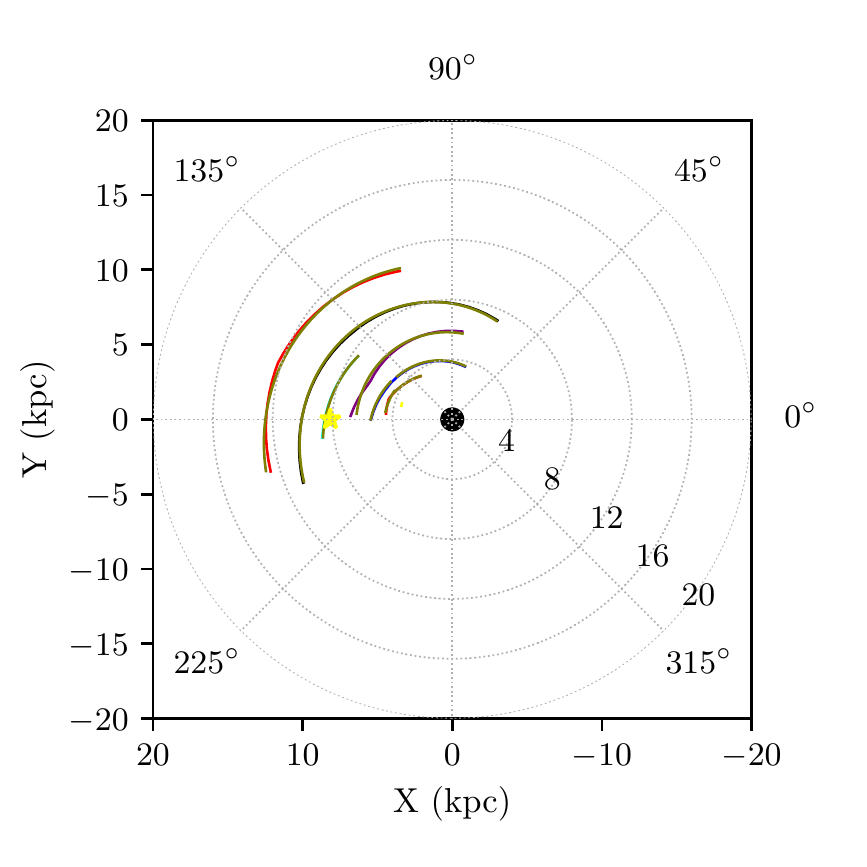}
  \caption{Retrieved spiral segments (olive) over-plotted on the original spiral arms delineated by \citet{Reid2019} used to define the mock sample of stars. The color-coding of the spiral arms is the same as in the original paper.}
  \label{fig:App1b}
\end{figure*}

 Fig.~\ref{fig:App1a} displays the original mock data (N$_{spiral}$=1500 Cepheids) in the Milky Way plane and in the \texttt{t-SNE} space. Given its small angular extension in the \citet{Reid2019} model, the 3-kpc arm is usually populated with only a handful of Cepheids (or less) and will be ignored in what follows. In all the tests, we did not try to adjust the hyper-parameters of \texttt{t-SNE+HDBSCAN} to the specifics of the mock data. and we kept the same values as for the real data. The very high values for the p-values and the recovery fractions in Table~\ref{tab:retrieved_params_case1} certify that the spiral arms are almost perfectly recovered by the algorithm. This can be seen as well in Fig.~\ref{fig:App1b}. The Norma and the Sct-Cen arms are recovered via two segments resulting from under-densities or gaps in the spatial distribution of Cepheids.\\
Increasing the number of stars in the mock catalog reduces the number and extension of such gaps, leading to the recovery of mock spiral arms within a single structure. In Fig.~\ref{fig:App1a}, we note that the other arms (e.g., the Perseus arm) also show gaps in the distribution of mock Cepheids, without being split into segments. From this, we conclude that segments are more likely to occur in crowded regions of the \texttt{t-SNE} space.

\begin{table}[ht]
\centering
\caption{Groups retrieved by the algorithm that correspond to a given spiral arm.}
\footnotesize
\label{tab:retrieved_params_case1}
\begin{tabular}{rrrrrr}
\hline\hline
     Arm &   Group & p-value & N$_{spiral}$ & Retrieved & \% retrieved\\
\hline
   Norma &  [7, 6] &    0.98 &           93 &        92 &        98.92 \\
 Sct-Cen &  [4, 5] &    1.00 &          169 &       168 &        99.41 \\
 Sgr-Car &     [0] &    1.00 &          167 &       167 &       100.00 \\
   Local &     [2] &    1.00 &          158 &       158 &       100.00 \\
 Perseus &     [3] &    1.00 &          550 &       550 &       100.00 \\
   Outer &     [1] &    1.00 &          362 &       362 &       100.00 \\
\hline
\end{tabular}
\tablefoot{The number and percentage of retrieved stars compared to the original number N$_{spiral}$ of mock stars in the spiral arm is provided, as well as the p-value testing the null hypothesis that the mock and retrieved spiral arms are identical using Hotelling’s t-Squared statistics.}
\end{table}

\FloatBarrier
\subsection{Test case 2: Noisy spiral arms}

\par In this test case, we increase the dispersion of mock Cepheids in spiral arms by adding random noise to their (x,y) coordinates. To achieve this, we added random values to them, which were sampled from a univariate standard normal distribution scaled by the "noise" parameter (see Table~\ref{tab:2D-MVN_params}). In this example, the noise parameter is 0.03.   
    
\par Figs.~\ref{fig:App2a} and \ref{fig:App2b} display the outcome of this test performed on a sample of N$_{spiral}$=1500 Cepheids. The algorithm performs similarly well as in the ideal case, and the mock spiral arms are very well reproduced. However, despite a recovery rate approaching 95\%, the p-value for the Sct-Cen strongly drops. A careful inspection of Fig.~\ref{fig:App2a} indicates that two stars attributed to group 8 rather belong to group 7. This misclassification leads to the inclusion of group 8 in the comparison to the original mock data for the Sct-Cen, and hence to the low p-value. We also note that a few stars close to the gap in the Sct-Cen arm remain unclassified (probably due to their proximity to the Norma arm), 

\begin{table}[ht]
\centering
\caption{Same as Table~\ref{tab:retrieved_params_case1}.}
\footnotesize
\label{tab:retrieved_params_case2}
\begin{tabular}{llrrrrrr}
\hline\hline
     Arm &         Group & p-value & N$_{spiral}$ & Retrieved & \% retrieved \\
\hline
   Norma &        [4, 8] &    0.94 &           92 &        91 &        98.91 \\
 Sct-Cen &  [5, 6, 7, 8] &    0.01 &          150 &       142 &        94.67   \\
 Sgr-Car &           [1] &    1.00 &          159 &       159 &       100.00   \\
   Local &           [2] &    1.00 &          161 &       161 &       100.00 \\
 Perseus &           [3] &    1.00 &          553 &       553 &       100.00   \\
   Outer &           [0] &    1.00 &          384 &       384 &       100.00 \\
\hline
\end{tabular}
\tablefoot{In this test case, random noise is added to the (x,y) coordinates of the stars in the mock spiral arms.}
\end{table}

\par
\begin{figure}[ht]
  \includegraphics[width=\linewidth]{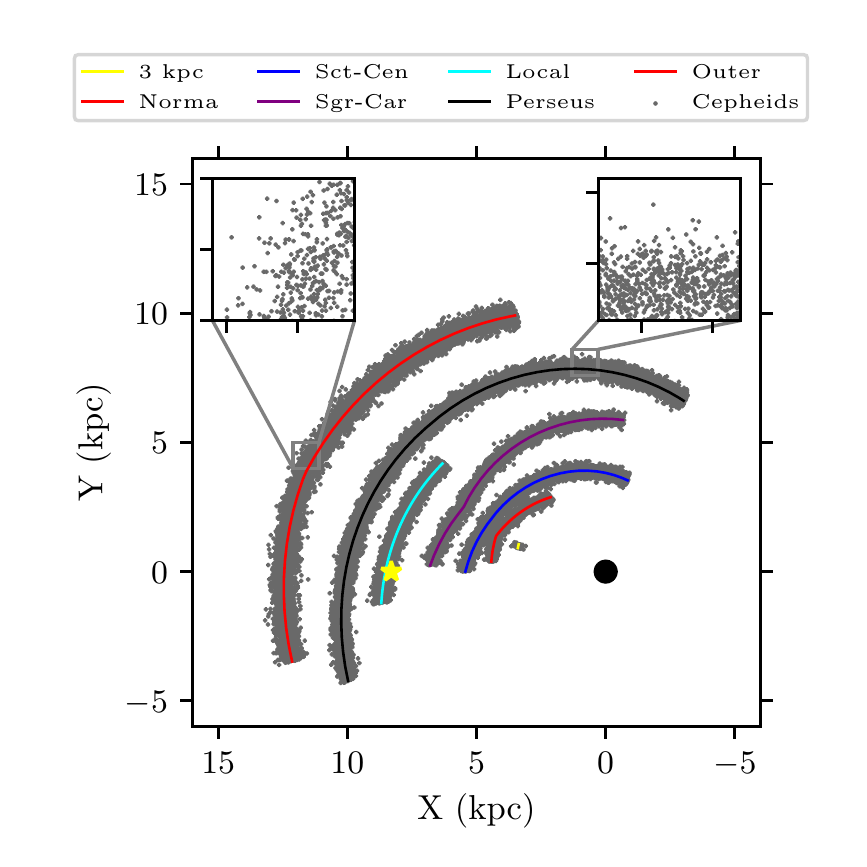}
  \caption{Full sample of mock Cepheids in spiral arms following the model by \citet{Reid2019}. The noise parameter in this example is 0.02. Cepheids created outside of the angular range covered by a given spiral arm have been discarded. From this sample, N$_{spiral}$ mock Cepheids are randomly selected.}
  \label{fig:Simulated_large_noise}
\end{figure}

\begin{figure*}[!hb]
  \includegraphics[width=\linewidth]{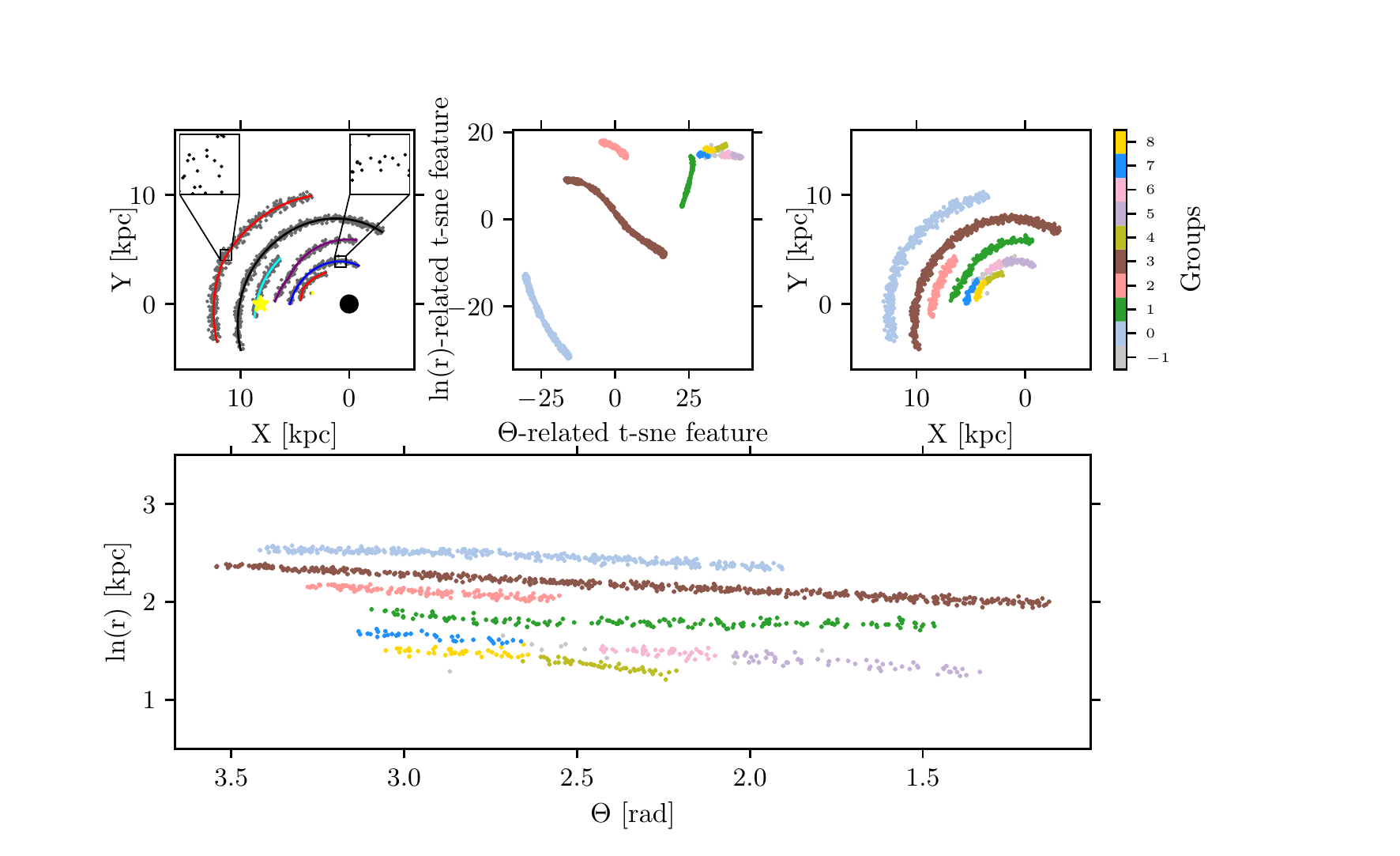}
  \caption{Same as Fig.~\ref{fig:App1a}, but with random noise added to the (x,y) coordinates of the stars in the mock spiral arms.}
  \label{fig:App2a}
\end{figure*}

\begin{figure*}[ht]
  \sidecaption
  \includegraphics[width=0.5\linewidth]{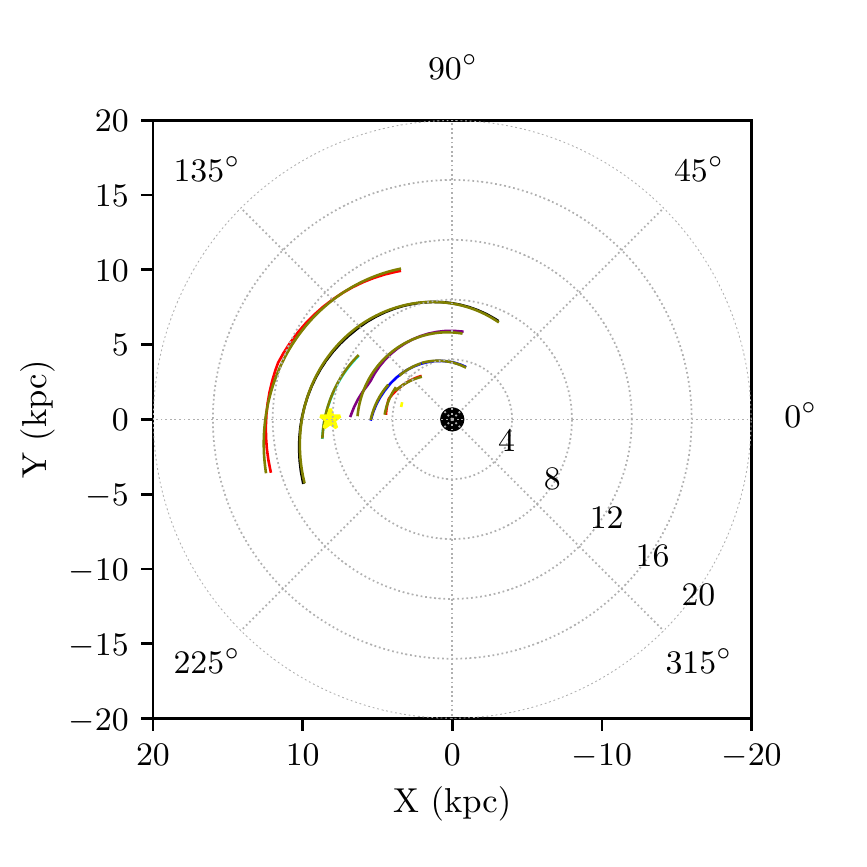}
  \caption{Same as Fig.~\ref{fig:App1b}.}
  \label{fig:App2b}
\end{figure*}

 This example remains a bit unrealistic because mock Cepheids stay confined to the spiral arms, and because the proximity of the Norma and Sct-Cen spiral arms prevents us from increasing the dispersion of individual spiral arms to very large values: Fig.~\ref{fig:Simulated_large_noise} shows that the mock Norma and Sct-Cen spiral arms come close to touching each other already with a noise parameter equals to 0.02.

\clearpage
\FloatBarrier

\clearpage
\FloatBarrier

\subsection{Test case 3: Inter-arms' Cepheids}

\begin{minipage}[b]{0.49\textwidth}
To provide a more realistic test case, we add to the mock sample of Cepheids in spiral arms (N$_{spiral}$=900, noise=0.01) a large collection of N$_{other}$=1500 Cepheids, whose coordinates are drawn from a bivariate normal distribution with a mean centered on the Sun and a variance of 80. These Cepheids can be considered as inter-arm Cepheids, although by construction some of them may overlap the spiral arms or be located in their immediate proximity.\\

In this example, almost all spiral arm members are retrieved, but the \texttt{t-SNE} groups also wrongly include $\approx$15\% to $\approx$55\% nonmembers, leading to lower p-values, and, in the case of the Norma and Sct-Cen arm, to a partial mismatch between the input and retrieved spiral arms (also leading to low p-values for these two arms). Indeed, Fig.~\ref{fig:App3b} shows that in the region where they are closest to each other, Norma and Sct-Cen are merged into a single structure. Fig.~\ref{fig:App3b} also shows that the other spiral arms are well retrieved, sometimes by the means of several segments. An artificial structure (group 9) emerges from the noise because it is relatively isolated in the \texttt{t-SNE} space and could have been falsely identified as a real structure beyond the outer arm.
\vspace{1cm}
\end{minipage}%
\begin{minipage}[b]{0.5\textwidth}
%\begin{figure}[!ht]
  \includegraphics[width=1.0\textwidth]{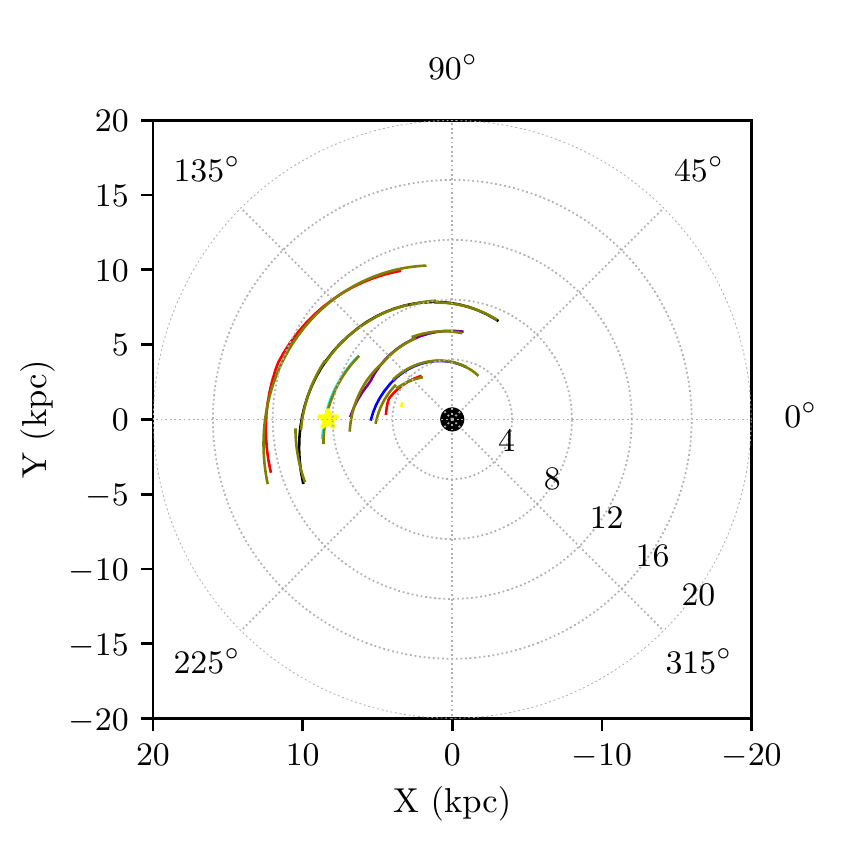}
  \captionof{figure}{Same as Fig.~\ref{fig:App1b}.}
  \label{fig:App3b}
%\end{figure}  
\end{minipage}

\begin{table*}[!b]
\centering
\caption{Same as Table~\ref{tab:retrieved_params_case1}.}
\label{tab:retrieved_params}
\begin{tabular}{llrrrrrr}
\hline\hline
     Arm &                     Group & p-value & N$_{spiral}$ & Retrieved & \% retrieved &  Extra & \% extra \\
\hline
   Norma &                    [7, 4] &    0.00 &           56 &        56 &       100.00 &  11 &    19.64 \\
 Sct-Cen &                    [8, 4] &    0.47 &          108 &       108 &       100.00 &  25 &    23.15 \\
 Sgr-Car &                 [2, 6, 5] &    0.02 &           98 &        98 &       100.00 &  35 &    35.71 \\
   Local &                       [0] &    0.27 &           88 &        88 &       100.00 &  24 &    27.27 \\
 Perseus &  [14, 16, 15, 13, 12, 17] &    0.43 &          341 &       337 &        98.83 &  49 &    14.37 \\
   Outer &                       [1] &    0.00 &          208 &       208 &       100.00 & 113 &    54.33 \\
\hline
\end{tabular}
\tablefoot{In this test case, random noise is added to the (x,y) coordinates of the stars in the mock spiral arms, and a large number of inter-arms' Cepheids is included in the sample. The table also indicates the number and percentage of inter-arms' stars wrongly included in the spiral arms.}
\end{table*}

\begin{figure*}[!b]
  \includegraphics[width=\linewidth]{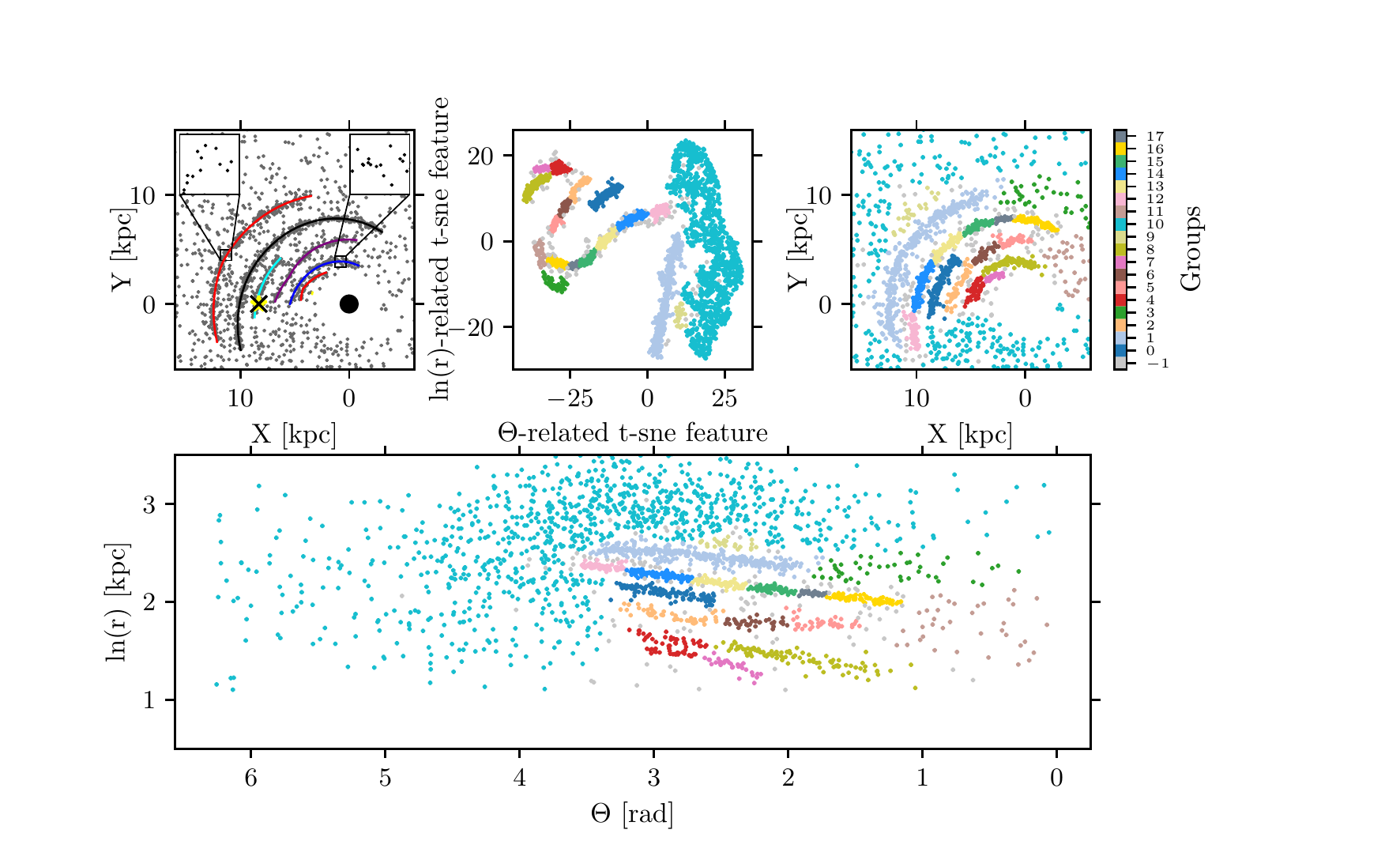}
  \caption{Same as Fig.~\ref{fig:App1a}, but with random noise added to the (x,y) coordinates of the $N_{spiral}$=900 stars in the mock spiral arms, and N$_{other}$=1500 inter-arms' Cepheids.}
  \label{fig:App3a}
\end{figure*}

%\par To provide a more realistic test case, we add to the mock sample of Cepheids in spiral arms (N$_{spiral}$=900, noise=0.01) a large collection of N$_{other}$=1500 Cepheids, whose coordinates are drawn from a bivariate normal distribution with a mean centered on the Sun and a variance of 80. These Cepheids can be considered as inter-arm Cepheids, although by construction some of them may overlap the spiral arms or be located in their immediate proximity.

%\par  In this example, almost all spiral arm members are retrieved, but the \texttt{t-SNE} groups also wrongly include $\approx$15\% to $\approx$55\% nonmembers, leading to lower p-values, and, in the case of the Norma and Sct-Cen arm, to a partial mismatch between the input and retrieved spiral arms (also leading to low p-values for these two arms). Indeed, Fig.~\ref{fig:App3b} shows that in the region where they are closest to each other, Norma and Sct-Cen are merged into a single structure. Fig.~\ref{fig:App3b} also shows that the other spiral arms are well retrieved, sometimes by the means of several segments. An artificial structure (group 9) emerges from the noise because it is relatively isolated in the \texttt{t-SNE} space and could have been falsely identified as a real structure beyond the outer arm.

%\begin{figure}[!ht]
%  \includegraphics[width=\linewidth]{App3b_simulated_arms_sig_0.1_noise_0.01_Gaussian_900_pts_1500_noise_segments.pdf}
%  \caption{Same as Fig.~\ref{fig:App1b}.}
%  \label{fig:App3b}
%\end{figure}  

\newpage
\onecolumn
\FloatBarrier

\section{Characterization of spiral arms} 
\label{App:spiral_age_cuts}

\begin{figure*}[!ht]
    \centering
    \includegraphics[width=\linewidth]{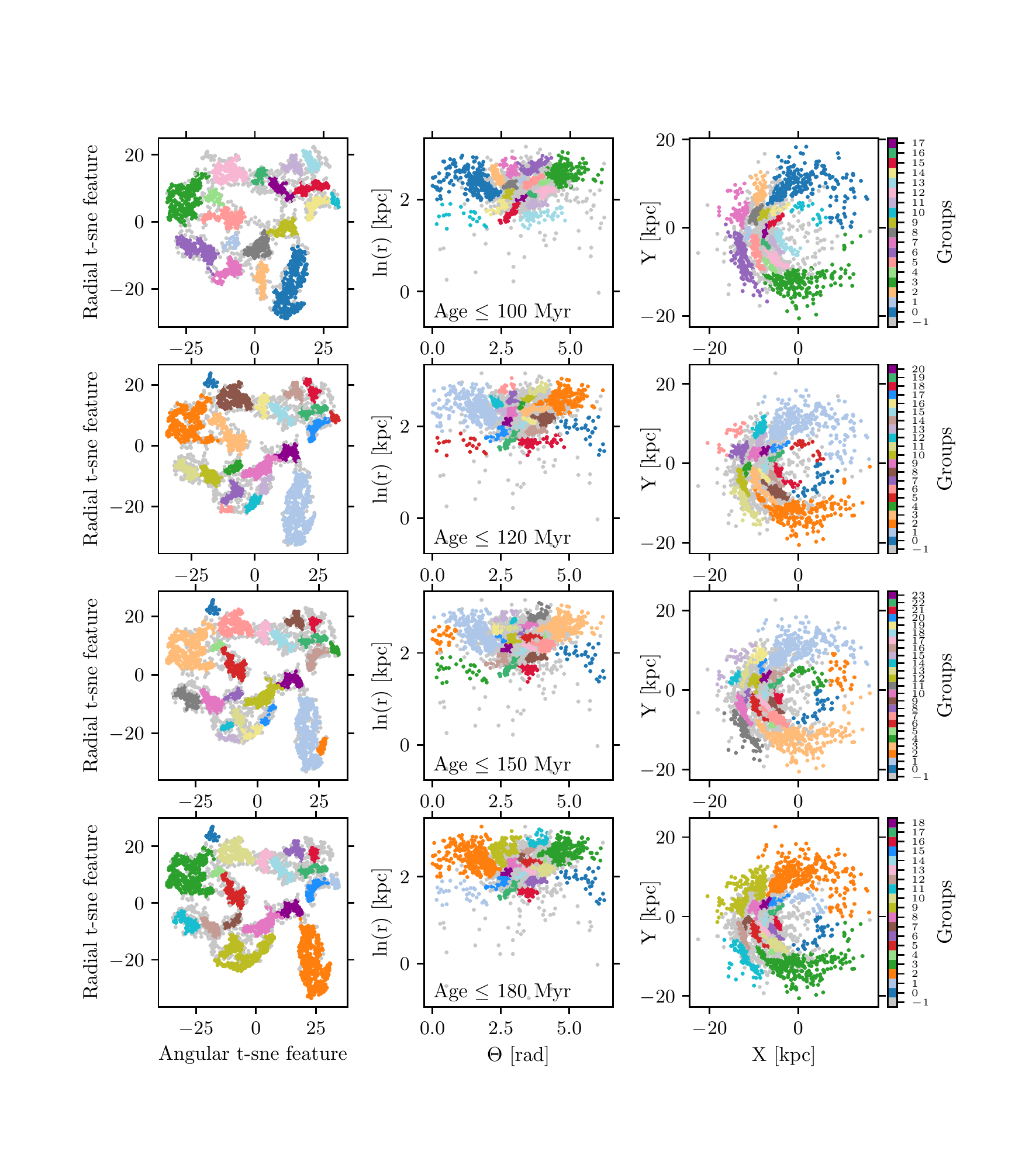}
    \caption{Outcome of the algorithm for different age cuts. The left panels show groups of Cepheids identified via \texttt{HDBSCAN} in the \texttt{t-SNE} space. These groups are then displayed in the ($\theta, ln r$) plane (middle panels), where $\theta$ is the Galactocentric azimuth and ln\,$r$ the logarithm of the Galactocentric radius (corrected from the warp), and in the Galactic plane (right panels), where the Galactic center is located at (0,0). The groups have been identified in different subsamples of the original Cepheid catalog by applying different age cuts, from Cepheids younger than 100\,Myr (top row) to Cepheids younger than 180\,Myr (bottom row).}
    \label{fig:age_cuts}
\end{figure*}

\begin{figure*}[ht]
    \centering
    \includegraphics[width=\linewidth]{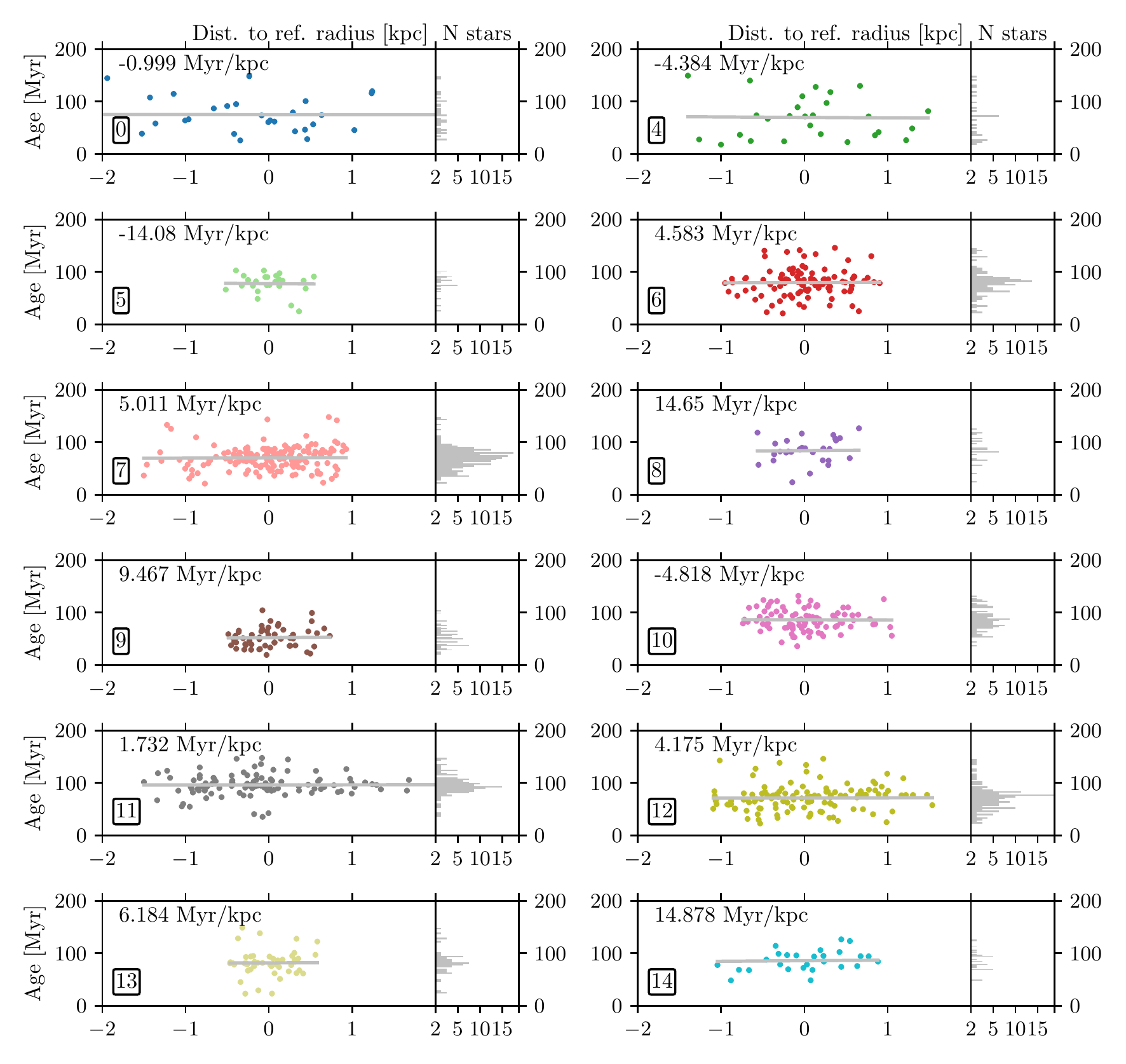}
    \caption{Age distribution of Cepheids in individual spiral segments or groups of segments. The age (in megayears) is plotted against the distance (in kiloparsecs) to the reference radius of the segment. A linear fit to the data indicates the age gradient across the considered segment (the slope of the gradient is provided). As in Figs.~\ref{fig:single_age_150}--\ref{fig:azimuthal_gradient}, only Cepheids younger than 150\,Myr are considered.}
    \label{fig:age_gradient_a}
\end{figure*}
\newpage
\begin{figure*}[ht]\ContinuedFloat
    \centering
    \includegraphics[width=\linewidth]{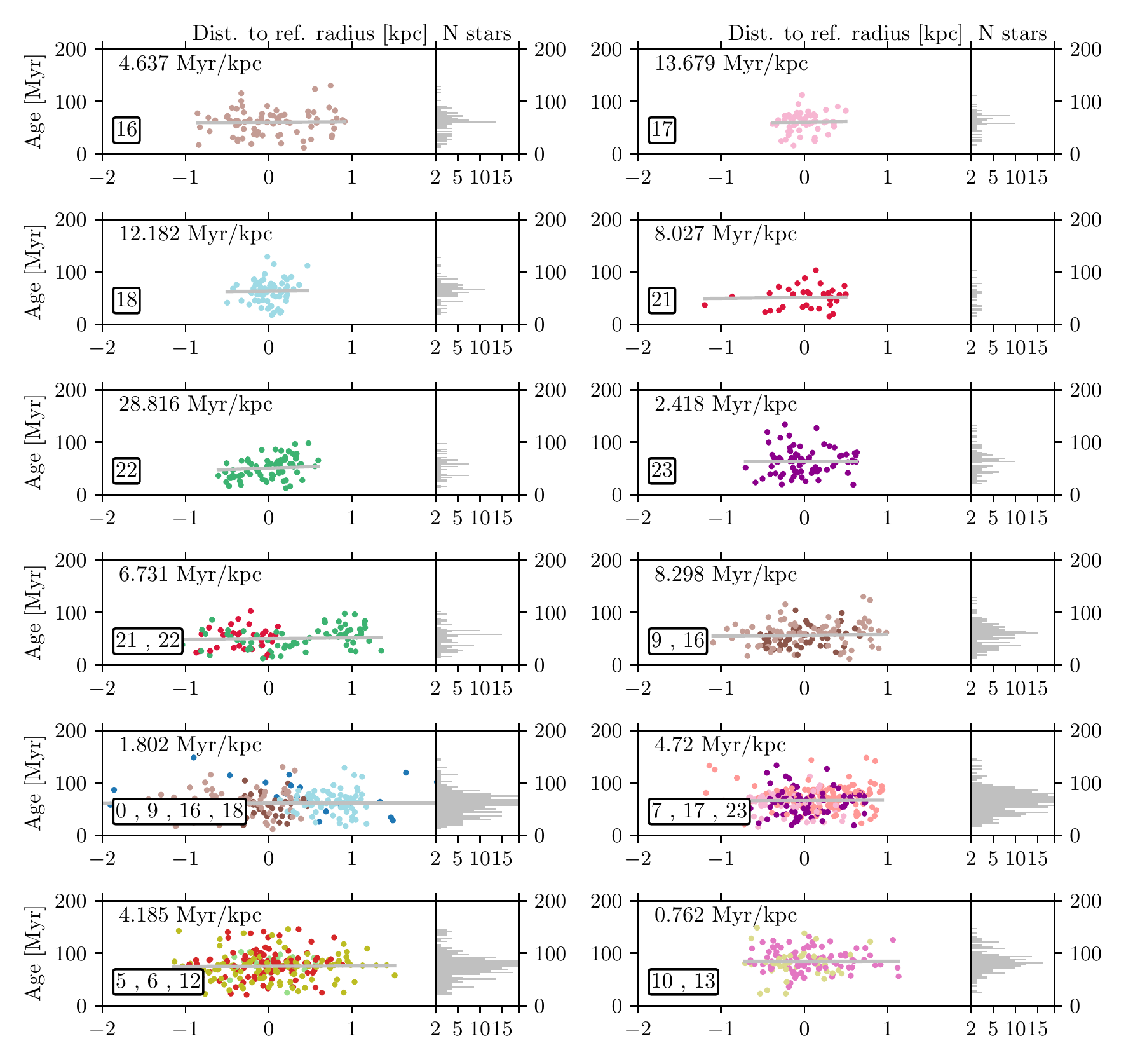}
    \caption{continued.}
    \label{fig:age_gradient_b}
\end{figure*}

\FloatBarrier
\section{Spiral segments: Alternative age cuts}
\label{App:spiral_arms}

\subsection{Cepheids younger than 100\,Myr}

\begin{figure*}[htpb]
    \centering
    \includegraphics[width=\linewidth]{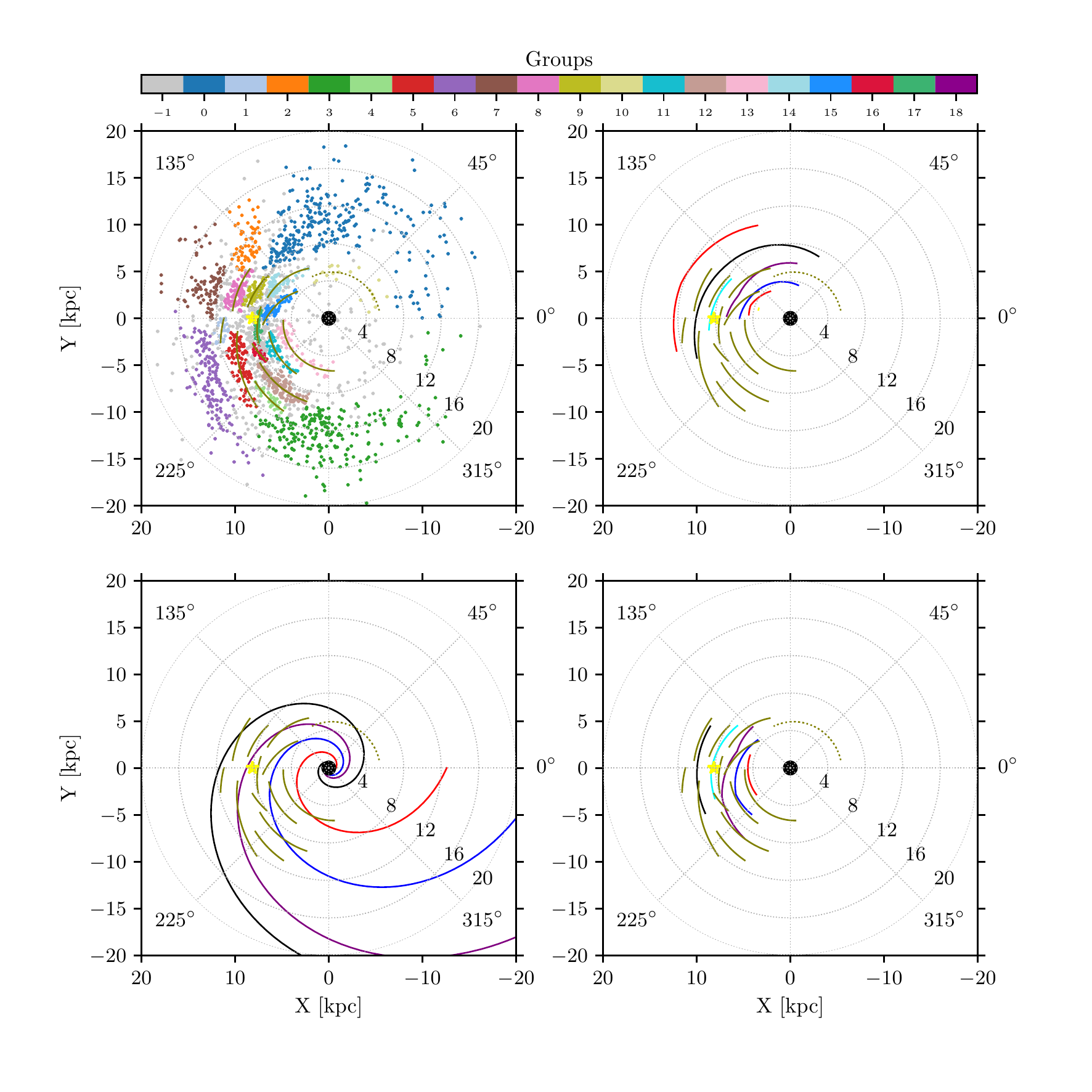}
    \caption{Same as Fig.~\ref{fig:plot_spiral_segments_150}, but for Cepheids younger than 100\,Myr.}
    \label{fig:plot_spiral_segments_100}
\end{figure*}

\begin{table*}[htpb]
\footnotesize
\centering
\caption{Same as Table~\ref{tab:params_segments_150}, but for Cepheids younger than 100\,Myr.}
\label{tab:params_segments_100}
\begin{tabular}{ccccccccccc}
\hline\hline
Group &  Slope & $\sigma_{slope}$ & Intercept & $\sigma_{intercept}$ & Ref. angle & Ref. radius & ln(Ref. radius) & Pitch angle & Min. angle & Max. angle \\
      & (kpc.rad$^{-1}$) & (kpc.rad$^{-1}$) &      (kpc) & (kpc) &    (rad) &      (kpc) &         (kpc) &        (rad) &     (rad) &     (rad) \\
\hline
 {[}10] & -0.071 &       0.060 &     1.706 &       0.073 &     1.053 &      5.110 &         1.631 &        0.071 &     0.154 &     1.951 \\
 {[}13] &  0.092 &       0.027 &     1.287 &       0.105 &     4.001 &      5.234 &         1.655 &       -0.092 &     3.190 &     4.812 \\
 {[}15] &  0.591 &       0.042 &     0.043 &       0.123 &     2.835 &      5.575 &         1.718 &       -0.534 &     2.436 &     3.233 \\
 {[}14] &  0.215 &       0.049 &     1.327 &       0.126 &     2.384 &      6.294 &         1.840 &       -0.212 &     1.961 &     2.807 \\
 {[}11] &  0.052 &       0.033 &     1.706 &       0.122 &     3.776 &      6.702 &         1.902 &       -0.052 &     3.375 &     4.177 \\
 {[}17] &  0.166 &       0.018 &     1.500 &       0.060 &     3.232 &      7.664 &         2.037 &       -0.164 &     2.980 &     3.485 \\
  {[}9] &  0.211 &       0.047 &     1.535 &       0.132 &     2.765 &      8.318 &         2.118 &       -0.208 &     2.534 &     2.996 \\
 {[}16] & -0.242 &       0.059 &     2.990 &       0.212 &     3.602 &      8.318 &         2.118 &        0.237 &     3.467 &     3.736 \\
 {[}12] &  0.065 &       0.025 &     1.927 &       0.100 &     4.084 &      8.957 &         2.192 &       -0.065 &     3.718 &     4.450 \\
  {[}8] &  0.073 &       0.042 &     2.108 &       0.120 &     2.821 &     10.114 &         2.314 &       -0.073 &     2.583 &     3.060 \\
  {[}4] &  0.148 &       0.045 &     1.768 &       0.183 &     4.055 &     10.677 &         2.368 &       -0.147 &     3.855 &     4.255 \\
  {[}5] &  0.283 &       0.028 &     1.356 &       0.102 &     3.656 &     10.921 &         2.391 &       -0.276 &     3.288 &     4.025 \\
  {[}1] &  0.252 &       0.086 &     1.624 &       0.281 &     3.251 &     11.510 &         2.443 &       -0.247 &     3.141 &     3.360 \\
\hline
\end{tabular}
\end{table*}

\FloatBarrier

\subsection{Cepheids younger than 120\,Myr}

\begin{table*}[hb]
\footnotesize
\centering
\caption{Same as Table~\ref{tab:params_segments_150}, but for Cepheids younger than 120\,Myr.}
\label{tab:params_segments_120}
\begin{tabular}{ccccccccccc}
\hline\hline
Group &  Slope & $\sigma_{slope}$ & Intercept & $\sigma_{intercept}$ & Ref. angle & Ref. radius & ln(Ref. radius) & Pitch angle & Min. angle & Max. angle \\
      & (kpc.rad$^{-1}$) & (kpc.rad$^{-1}$) &      (kpc) & (kpc) &    (rad) &      (kpc) &         (kpc) &        (rad) &     (rad) &     (rad) \\
\hline
  {[}5] & -0.039 &       0.051 &     1.650 &       0.063 &     1.053 &      4.998 &         1.609 &        0.039 &     0.154 &     1.951 \\
 {[}18] &  0.039 &       0.025 &     1.503 &       0.094 &     3.994 &      5.253 &         1.659 &       -0.039 &     3.177 &     4.812 \\
 {[}19] &  0.570 &       0.043 &     0.099 &       0.125 &     2.835 &      5.556 &         1.715 &       -0.518 &     2.436 &     3.233 \\
 {[}17] &  0.212 &       0.049 &     1.333 &       0.125 &     2.384 &      6.287 &         1.838 &       -0.209 &     1.961 &     2.807 \\
 {[}14] &  0.046 &       0.031 &     1.731 &       0.116 &     3.743 &      6.707 &         1.903 &       -0.046 &     3.309 &     4.177 \\
  {[}0] & -0.228 &       0.073 &     3.169 &       0.405 &     5.424 &      6.906 &         1.932 &        0.224 &     4.566 &     6.282 \\
 {[}15] &  0.157 &       0.018 &     1.533 &       0.059 &     3.260 &      7.728 &         2.045 &       -0.156 &     2.980 &     3.541 \\
 {[}20] &  0.134 &       0.042 &     1.740 &       0.118 &     2.770 &      8.258 &         2.111 &       -0.133 &     2.534 &     3.007 \\
 {[}16] & -0.206 &       0.026 &     2.859 &       0.093 &     3.527 &      8.436 &         2.133 &        0.203 &     3.317 &     3.736 \\
  {[}8] &  0.039 &       0.024 &     2.030 &       0.098 &     4.044 &      8.915 &         2.188 &       -0.039 &     3.639 &     4.450 \\
  {[}9] &  0.152 &       0.073 &     1.878 &       0.211 &     2.896 &     10.157 &         2.318 &       -0.151 &     2.733 &     3.060 \\
  {[}3] &  0.054 &       0.013 &     2.156 &       0.048 &     3.737 &     10.568 &         2.358 &       -0.054 &     3.220 &     4.255 \\
  {[}4] &  0.281 &       0.072 &     1.541 &       0.233 &     3.274 &     11.716 &         2.461 &       -0.274 &     3.141 &     3.406 \\
 {[}10] &  0.046 &       0.021 &     2.444 &       0.073 &     3.511 &     13.538 &         2.605 &       -0.046 &     3.237 &     3.785 \\
 {[}11] &  0.190 &       0.028 &     2.036 &       0.107 &     3.851 &     15.923 &         2.768 &       -0.188 &     3.498 &     4.205 \\
\hline
\end{tabular}
\end{table*}

\begin{figure*}[htpb]
    \centering
    \includegraphics[width=\linewidth]{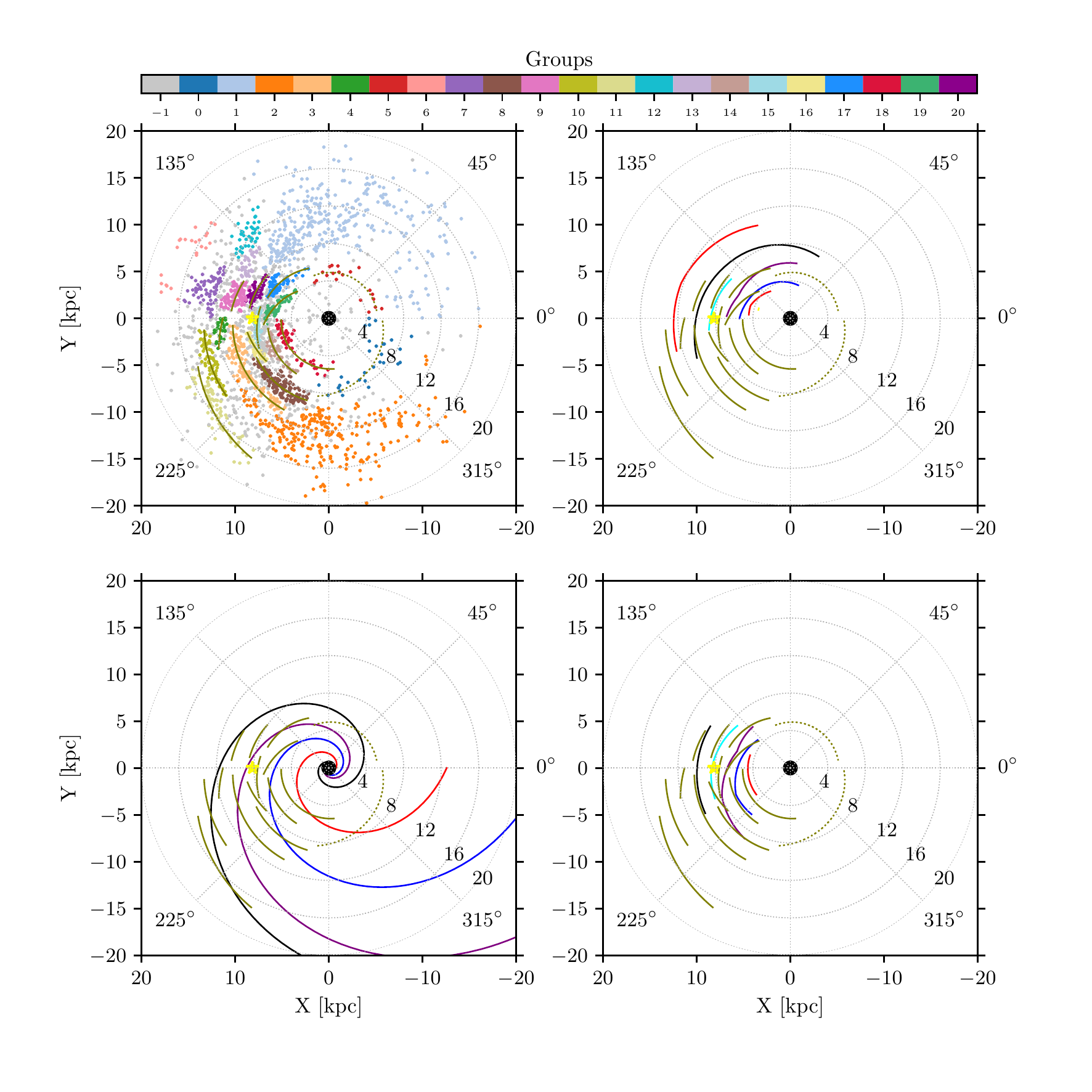}
    \caption{Same as Fig.~\ref{fig:plot_spiral_segments_150}, but for Cepheids younger than 120\,Myr.}
    \label{fig:plot_spiral_segments_120}
\end{figure*}

\FloatBarrier
\clearpage

\subsection{Cepheids younger than 180\,Myr}

\begin{figure*}[!hb]
    \centering
    \includegraphics[width=\linewidth]{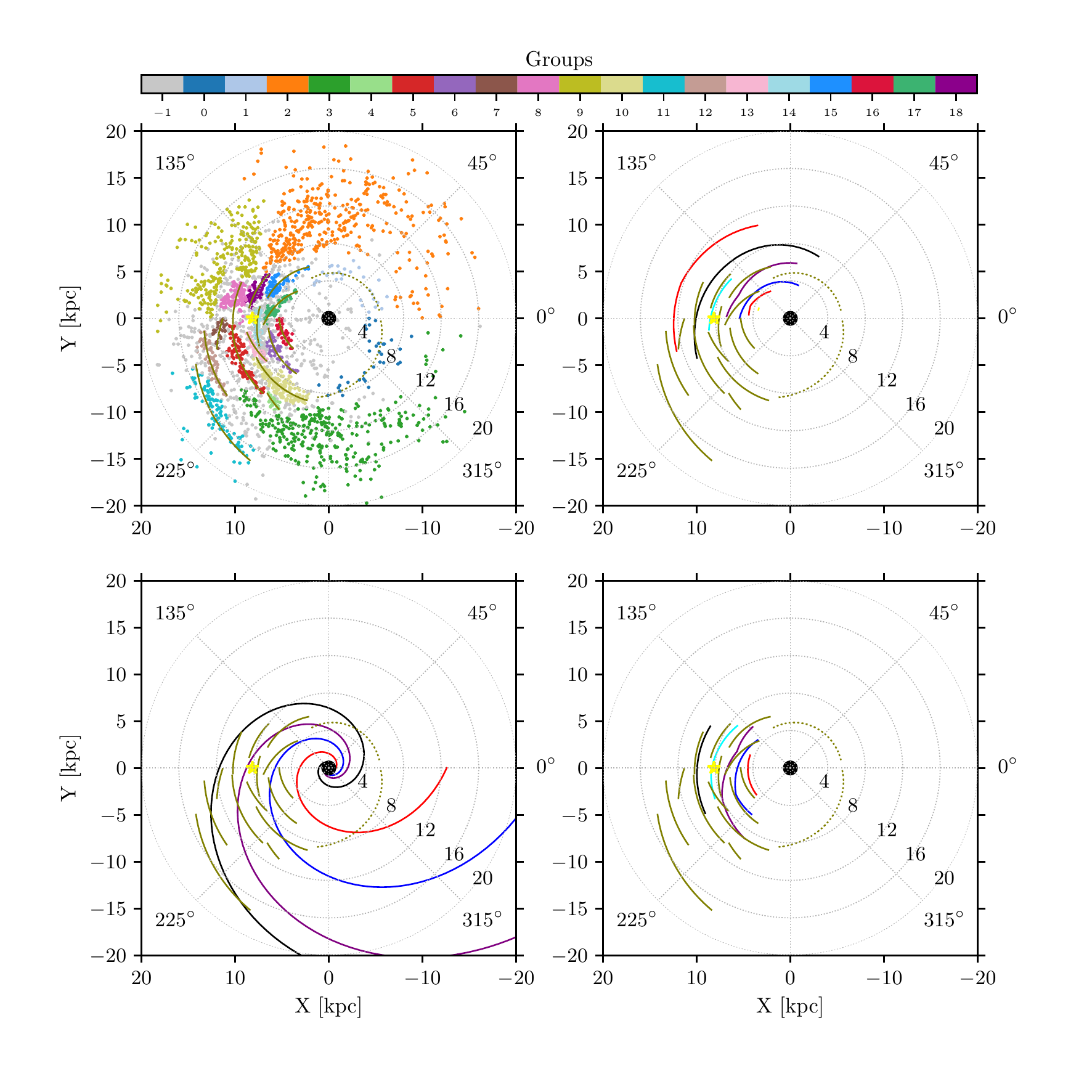}
    \caption{Same as Fig.~\ref{fig:plot_spiral_segments_150}, but for Cepheids younger than 180\,Myr.}
    \label{fig:plot_spiral_segments_180}
\end{figure*}

\begin{table*}
\footnotesize
\centering
\caption{Same as Table~\ref{tab:params_segments_150}, but for Cepheids younger than 180\,Myr.}
\label{tab:params_segments_180}
\begin{tabular}{ccccccccccc}
\hline\hline
Group &  Slope & $\sigma_{slope}$ & Intercept & $\sigma_{intercept}$ & Ref. angle & Ref. radius & ln(Ref. radius) & Pitch angle & Min. angle & Max. angle \\
      & (kpc.rad$^{-1}$) & (kpc.rad$^{-1}$) &      (kpc) & (kpc) &    (rad) &      (kpc) &         (kpc) &        (rad) &     (rad) &     (rad) \\
\hline
  {[}1] & -0.087 &       0.051 &     1.712 &       0.062 &     1.068 &      5.048 &         1.619 &        0.087 &     0.154 &     1.982 \\
 {[}16] & -0.088 &       0.058 &     1.944 &       0.201 &     3.486 &      5.141 &         1.637 &        0.088 &     3.145 &     3.827 \\
 {[}17] &  0.570 &       0.041 &     0.100 &       0.121 &     2.835 &      5.561 &         1.716 &       -0.518 &     2.436 &     3.233 \\
 {[}15] &  0.182 &       0.046 &     1.417 &       0.118 &     2.379 &      6.359 &         1.850 &       -0.180 &     1.950 &     2.807 \\
  {[}6] &  0.053 &       0.032 &     1.700 &       0.119 &     3.743 &      6.675 &         1.898 &       -0.053 &     3.309 &     4.177 \\
  {[}0] & -0.256 &       0.071 &     3.312 &       0.391 &     5.424 &      6.845 &         1.923 &        0.251 &     4.566 &     6.282 \\
 {[}14] &  0.128 &       0.018 &     1.627 &       0.059 &     3.253 &      7.717 &         2.043 &       -0.127 &     2.980 &     3.526 \\
 {[}18] &  0.148 &       0.039 &     1.703 &       0.108 &     2.759 &      8.259 &         2.111 &       -0.147 &     2.511 &     3.007 \\
 {[}13] & -0.227 &       0.028 &     2.934 &       0.101 &     3.527 &      8.444 &         2.133 &        0.223 &     3.317 &     3.736 \\
 {[}10] &  0.038 &       0.023 &     2.035 &       0.096 &     4.044 &      8.924 &         2.189 &       -0.038 &     3.639 &     4.450 \\
  {[}8] &  0.038 &       0.064 &     2.205 &       0.186 &     2.977 &     10.157 &         2.318 &       -0.038 &     2.756 &     3.198 \\
  {[}5] &  0.038 &       0.020 &     2.214 &       0.072 &     3.602 &     10.495 &         2.351 &       -0.038 &     3.220 &     3.984 \\
  {[}4] &  0.357 &       0.124 &     0.900 &       0.511 &     4.120 &     10.705 &         2.371 &       -0.343 &     4.031 &     4.208 \\
  {[}7] &  0.344 &       0.096 &     1.344 &       0.313 &     3.278 &     11.841 &         2.472 &       -0.331 &     3.149 &     3.406 \\
 {[}12] &  0.037 &       0.020 &     2.472 &       0.072 &     3.517 &     13.492 &         2.602 &       -0.037 &     3.248 &     3.785 \\
 {[}11] &  0.197 &       0.038 &     2.024 &       0.146 &     3.842 &     16.134 &         2.781 &       -0.195 &     3.480 &     4.205 \\
\hline
\end{tabular}
\end{table*}

\end{appendix}

\end{document}